# Deconstructing Legal Text: Object-Oriented Design in Legal Adjudication

MEGAN MA*
DMITRIY PODKOPAEV**
AVALON CAMPBELL-COUSINS***
ADAM NICHOLAS****

## *Abstract*

Rules are pervasive in the law. In the context of computer engineering, the translation of legal text to algorithmic form is seemingly direct. In large part, law may be a ripe field for expert systems and machine learning. For engineers, existing law appears formulaic and logically reducible to 'if, then' statements. The underlying assumption is that the legal language is both self-referential and universal. Moreover, description is considered distinct from interpretation; that in describing the law, the language is seen as quantitative and objectifiable. Nevertheless, is descriptive formal language purely dissociative? From the logic machine of the 1970s to the modern fervor for artificial intelligence (AI), governance by numbers is making a persuasive return. Could translation be possible?

The project follows a fundamentally semantic conundrum: *what is the significance of 'meaning' in legal language?* The project, therefore, tests translation by deconstructing sentences from existing legal judgments to their constituent factors. Definitions are then extracted in accordance with the interpretations of the judges. The intent is to build an expert system predicated on alleged rules of legal reasoning. The authors apply both linguistic modelling and natural language processing technology to parse the legal judgments. The project extends beyond prior research in the area, combining a broadly statistical model of context with the relative precision of syntactic structure. The preliminary hypothesis is that, by analyzing the components of legal language with a variety of techniques, we can begin to translate law to numerical form.

* PhD Candidate in Law and Lecturer, Sciences Po.

** Senior Legal Data Scientist, Simmons Wavelength Ltd.

*** MASt Mathematics, University of Cambridge.

**** BA Linguistics, University of Cambridge.

## PREAMBLE

Rules are pervasive in the law. In the context of computer engineering, the translation of legal text to algorithmic form is seemingly direct. In large part, law may be a ripe field for expert systems and machine learning. For engineers, existing law appears formulaic and logically reducible to 'if, then' statements. The underlying assumption is that the legal language is both self-referential and universal. Moreover, description is considered distinct from interpretation; that in describing the law, the language is seen as quantitative and objectifiable. Nevertheless, is descriptive formal language purely dissociative? From the logic machine of the 1970s to the modern fervor for artificial intelligence (AI), governance by numbers is making a persuasive return. Could translation be possible?

Most recently, Douglas Hofstadter commented on the "Shallowness of Google Translate."[1] He referred largely to the Chinese Room Argument;[2] that machine translation, while comprehensive, lacked understanding. Perhaps he probed at a more important question: does translation *require* understanding? Hofstadter's experiments indeed seemed to prove it so. He argued that the purpose of language was not about the processing of texts. Instead, translation required imagining and remembering; "a lifetime of experience and [...] of using words in a meaningful way, to realize how devoid of content all the words thrown onto the screen by Google translate are."[3] Hofstadter describes the appearance of understanding language; that the software was merely 'bypassing or circumventing' the act.[4]

Yulia Frumer, a historian of science, notes that translation not only requires producing the adequate language of foreign ideas, but also the "situating of those ideas in a different conceptual world."[5] That is, with languages that belong to the same semantic field, the conceptual transfer in the translation process is assumed. However, with languages that do not share similar intellectual legacies, the meaning of words must be articulated through the conceptual world in which the language is seated.

Frumer uses the example of 18th century Japanese translations of Dutch scientific texts. The process by which translation occurred involved first analogizing from Western to Chinese natural philosophy; effectively reconfiguring the foreign to local through experiential learning. This is particularly fascinating, provided that scientific knowledge inherits the reputation of universality. Yet, Frumer notes, "...if we attach meanings to statements by abstracting previous experience, we must acquire new experiences in order to make space for new interpretations."[6]

Mireille Hildebrandt teases this premise by addressing the inherent challenge of translation in the computer 'code-ification' process. Pairing speech-act theory with the mathematical theory of

1 Douglas Hofstadter, *The Shallowness of Google Translate*, The Atlantic (January 30, 2018), https://www.theatlantic.com/technology/archive/2018/01/the-shallowness-of-google-translate/551570/.

2 A thought-experiment first published by John Searle in 1980 arguing that syntactic rule-following is not equivalent to understanding.

3 Hofstadter, *supra* 1.

4 *Id.*

5 Yulia Frumer, *Translating Worlds, Building Worlds: Meteorology in Japanese, Dutch, and Chinese*, 109 ISIS 326 (2018).

6 *Id.* at 327.

information, she investigates the performativity of the law when applied to computing systems. In her analytical synthesis of these theories, she dwells on meaning. "Meaning," she states, "...depends on the curious entanglement of self-reflection, rational discourse and emotional awareness that hinges on the opacity of our dynamic and large inaccessible unconscious. Data, code...do not attribute meaning."[7] The inability of computing systems to process meaning raises challenges for legal practitioners and scholars. Hildebrandt suggests that the shift to computation necessitates a shift from reason to statistics. Learning to "speak the language" of statistics and machine-learning algorithms would become important in the reasoning and understanding of biases inherent in legal technologies.[8]

More importantly, the migration from descriptive natural language to numerical representation runs the risk of slippage as ideas are (literally) 'lost in translation.' Legal concepts must necessarily be reconceptualized for meaning to exist in the mathematical sense. The closest in semantic ancestry would be legal formalism. Legal formalists thrive on interpreting law as rationally determinate. Judgments are deduced from logical premises; meaning is assigned. While, arguably, the formalization of law occurs 'naturally' – as cases with like factual circumstances often form rules, principles, and axioms for treatment – the act of conceptualizing the law as binary and static is puzzling. Could the law behave like mathematics; and thereby the rule of law be understood as numeric?

Technology not only requires rules to be defined from the start, but that such rules are derived from specified outcomes. Currently, even with rules that define end-states, particularized judgments remain accessible. Machines, on the other hand, are built on logic and fixed such that the execution of tasks becomes automatic. Outcomes are characterized by their reproductive accuracy. Judgments, on the other hand, are rarely defined by accuracy. Instead, they are weighed against social consensus.

To translate the rule of law in a mathematical sense would require a reconfiguration of legal concepts. Interestingly, the use of statistics and so-called 'mathematisation' of law is not novel. Oliver Wendell Holmes Jr. most famously stated in the *Path of Law* that "[f]or the rational study of the law, the blackletter man may be the man of the present, but the man of the future is the man of statistics and the master of economics."[9] Governance by numbers then realizes the desire for determinacy; the optimization of law to its final state of stability, predictability, and accuracy. The use of formal logic for governance has a rich ancestry. The common denominator was that mathematical precision should be applied across all disciplines.

Since the twelfth century, mathematical logicians have used logical paradoxes to spot 'false' arguments in courts of law.[10] In the seventeenth century, Gottfried Leibniz proposed a mental alphabet;[11] whereby thoughts could be represented as combinations of symbols, and reasoning could be performed using statistical analysis. From Leibniz, George Boole's infamous treatise, *The Laws of Thought*,[12] argued that algebra was a symbolic language capable of expression and construction of argument. By the end of the twentieth century, mathematical equations were conceivably dialogic; a form of discourse.

[7] Mireille Hildebrandt, *Law* as *computation in the era of artificial intelligence: Speaking law to the power of statistics*, Draft for SPECIAL ISSUE U. TORONTO L.J. 10 (2019).

[8] Advances in natural language processing (NLP), for example, have opened the possibility of 'performing' calculations on words. This technology has been increasingly applied in the legal realm. See *id.* at 13.

[9] Oliver Wendell Holmes Jr., *The Path of Law*, 10 HARV. L. REV. 457, 469 (1897).

[10] Keith Devlin, *Goodbye Descartes: The End of Logic and The Search for a New Cosmology of the Mind* 54 (1997).

[11] *Id.* at 62.

This was perceivably owed to Boole's system; that complex thought could be reducible to the solution of equations. Nevertheless, the most fundamental contribution of Boole's work was the capacity to isolate notation from meaning.[13] That is, 'complexities' of the world would fall into the background as pure abstraction is brought to center stage. Eventually, Boole's work would form the basis of the modern-day algorithm and expression in formal language.

With the rise of artificial legal intelligence, computable law is making a powerful return. Legal texts may be represented as computational data with terms made 'machine-readable' through a process of conversion. Despite the capacity to express legal language in an alternative computable form, the notion of interpretation appears to have changed. How would digital data inscription and processing alter methods of legal reasoning?

### a. *Research Question and Outline of Approach*

The project follows a fundamentally semantic conundrum: what is the significance of 'meaning' in legal language? From a statistics standpoint, meaning can be approximated. Applying word analogies as the 'mathematical' basis, meaning is gauged by the statistical probability of the response. In recognizing the context and relationship between words, meaning hinges on the frequency of its appearance in a particular setup. That is, what do its neighbors reveal about the word in question?

Reflecting on Hildebrandt and Frumer, meaning is associated with experience; thereby finding meaning to legal concepts would require abstracting from experience. Should experience be built from conceptual worlds, to move across these worlds would be to translate. Translating legal language then requires a reframing of legal concepts; perhaps an expression of the law based on statistical experience as opposed to natural language.

The project will proceed in two phases: (1) the proof of concept (POC); and (2) application to broader legal corpora. In the first phase, the POC will analyze three United States Supreme Court cases. The selection was chosen on the basis of a similar factual premise and time frame. That is, all three cases involve defining the use of firearms and were ruled in rapid succession. These cases are *Smith v. United States* (1993), *Bailey v. United States* (1995), and *Muscarello v. United States* (1998). While there are evidently a number of caveats[14] to this selection, it nonetheless has merit as an interesting starting point. Notably, the POC wrestles with the existence of legal concepts. The goals of the POC are two-fold: (1) to analyze the processes involved with legal interpretation and reasoning; and (2) critically assess them against the function of law.

[12] George Boole, *The Laws of Thought* Chapter 1 (1854).

[13] *Id.* at 77.

[14] Some of these caveats include selection bias, sample size, and perhaps more importantly, an amendment has since been made to the legislation in question.

Methodologically, the POC tests translation by deconstructing sentences from existing legal judgments to their constituent factors. Definitions are then extracted in accordance with the interpretations of the judges. The intent is to build an expert system predicated on alleged rules of legal reasoning. The authors intend to apply both linguistic modelling and natural language processing (NLP) technology to parse the legal judgments. The preliminary hypothesis is that, by analyzing the components of legal language with a variety of techniques, we can begin to translate law to numerical form. Furthermore, it would be interesting to consider what contextual understanding may need to exist to understand the language of various legal documents.

Following the POC, the authors will extend the test to the corpora of United States Supreme Court cases. This stage of the research will consider the feasibility of expanding the approach to similar legal texts. For the purposes of the current paper, the authors will focus on the observations and findings from the POC. The authors believe that the initial assessment at the POC stage could contribute to a more fruitful dialogue on the integration of computational technology in law.

The POC falls in line with existing literature on Law2Vec and legal word embeddings. Equally, the project extends beyond prior research in the area, combining a broadly statistical model of context with the relative precision of syntactic structure. In effect, the POC intends to generate building blocks to determine "context" explained in the text; thereby able to define the use of firearms through a framework of extraction.

The paper will proceed as follows. Part I will begin with a literature review of texts that have fueled the project's inquiries and formed the environment which it intends to resolve. As the nature of the paper is fundamentally interdisciplinary, it draws reference from law, linguistics, and computer science. Part II discusses the methodology the authors have taken; highlighting both elements of inspiration and strategies considered. Part III teases at preliminary observations and notes of interest during the project's progression. Part IV details the technological implementation and the actual steps towards translation. Part V reflects on early achievements and areas of further advancement. The authors will then conclude with final and next steps.

## I. LITERATURE REVIEW

### *a. Jurisprudential Heritage*

AI adjudication is an evidently polarized subject. Questions around the prospect of "robot judges" typically center on morality and equitable justice;[15] on issues of explanability and Black Box machine learning.[16] In common law systems, the art of drafting legal opinions begins with mastering legal argumentation. To ground the argument within the sphere of existing legal texts is the linchpin of judicial decisions.

Legal theory becomes a referencing point when courts are asked to interpret legal documents. Textualism, for example, "narrow[s] the range of acceptable judicial decision-making and acceptable argumentation"[17] by turning to dictionary definitions and rejecting judicial speculation. Yet, what is the purpose of 'narrowing the range'? To that question, Antonin Scalia answers, "...textualism will provide greater certainty in the law, and hence greater predictability..."[18] So, what are its assumptions and implications? Eric Posner suggests, there may be aspirational intentions "to keep the law pure";[19] or otherwise, to ensure that the legal system is consistent. Textualism also reinforces the role of judges. That is, judges are to interpret passively, and that legal interpretations are to be semantic.[20]

[15] Richard M. Re and Alicia Solow-Niederman, *Developing Artificially Intelligent Justice*, 22 STAN. TECH. L. REV. 242 (2019).

[16] *See* Yavar Bathaee, *The Artificial intelligence Black Box and the Failure of Intent and Causation*, 31 HARV. J OF L. & TECH 890; and also, Frank Pasquale, *Black Box Society: The Secret Algorithms that Control Money and Information* (2015).

Consider the infamous example of a municipal legislation stating that "no person may bring a vehicle into the park."[21] Would an ambulance be permitted to enter the park in the event of an accident? For textualists, they may argue that – according to the dictionary definition – an ambulance is a vehicle; and thereby, cannot enter the park. Should the legislators have thought an ambulance was an exception, they would have included it in the text. Accepting the premise of that argument, what about a police car or a firetruck? Perhaps the legislation should be amended to include all emergency vehicles. What happens then if an ambulance is merely parked inside the park with no foreseeable emergency?

The example illustrates that the problem with textualism becomes rapidly cyclical, as interpretations rendered must either become increasingly narrow or increasingly broad to accommodate a "myriad [of] hypothetical scenarios and provide for all of them explicitly."[22] Textualism, therefore, falls down the slippery slope of literalism. Words of legal texts are assumed to embody intrinsic meaning and are waiting to be extracted.

Moreover, the impact of mere 'extraction' is its precedential value. The approach, taken most prominently in common law systems, is to follow past decisions. Adopting the decisions of the past to guide future conduct parallels this exact act of extraction. That is, applying past precedents provides the scope for a "gradual moulding of the rules to meet fresh situations as they arise."[23] Decisions have binding legal force. Interpretations of the past should carry the definitions to be used moving forward. The role of the judge is that of an archaeologist; excavating legal truths from judicial past.

This is seemingly straightforward. Yet, the challenge encountered is identifying within the decision the kernel of precedent. Holmes describes the challenge as a paradox of form and substance in the development of the law. The form is logical, as "each new decision follows syllogistically from existing precedents."[24] Still, its substance is legislative and draws on views of public policy. Holmes argues that the law is driven by the "unconscious result of instinctive preferences and inarticulate convictions;" and therefore, "the law [is] always approaching, and never reaching, consistency."[25] The ostracized conclusion would be that judicial decisions have an element of inexplicability, and are, in fact, a 'Black Box.'[26] Recalling Hildebrandt, "meaning" becomes a metaphor and the heart of the juridical process.

[17] Antonin Scalia and Bryan A. Garner, *Reading Law: The Interpretation of Legal Texts* xxvii-xxix (2012).

[18] *Id.*

[19] Eric Posner and Adrien Vermeule, *Inside or Outside the System?*, 80 U. CHI. L. REV. 1743, 1775 (2013).

[20] Richard A. Posner, *The Incoherence of Antonin Scalia*, New Republic (August 24, 2012), http://www.newrepublic.com/node/106441/print.

[21] Taken originally from H.L.A. Hart where he posed the hypothetical of "no vehicles in the park." *See* H.L.A. Hart, *Positivism and the Separation of Law and Morals*, 71 HARV. L. REV. 593, 607 (1958). This has often been referenced in legal literature. *See* for example, Pierre Schlag, *No vehicles in the Park*, 23 SEATTLE U. L. REV. 381, 382 (1999); and more recently, Michael Genesereth, *Computational Law: The Cop in the Backseat*, White Paper, CodeX: The Center for Legal Informatics (2015), available at: http://logic.stanford.edu/publications/genesereth/complaw.pdf.

[22] Posner, *supra* 20.

[23] *See* chapter on Theories of Adjudication, in particular the discussion on *stare decisis* as the 'life blood of legal systems,' requiring precision in addition to stability and certainty. Michael Freeman, *Lloyd's Introduction to Jurisprudence* (9th ed., 2014).

The significance of the paper is, in part, to unpack the paradox articulated by Holmes. The selected cases aim to paint a picture on the use of precedent as a legal tool; and whether the law subconsciously follows a logic. To create the painting, the authors draw inspiration from the field of linguistics.

### *b. Linguistic Influence*

A grasp on the underlying hierarchical structure of language is key to breaking down sentences in a meaningful manner. Analyses of sentence structure fall primarily into two schools of thought: (1) dependency; and (2) phrase structure. The former, commonly represented as dependency trees, begins with the root verb of the superordinate clause and branches out from there, with subordinate verbs arranging substructures. Dependency trees map one node to each word without projecting constituent phrases: each word simply depends on another. For example, in most English sentences, the subject typically falls to the left of the verb, while its other dependencies (e.g. its objects) fall to the right. Since each word in a dependency syntax is represented by precisely one node, structural redundancy is arguably decreased. This system has been characterized as well-suited for algorithmic translation from natural language, owing to the node conservatism and predictability of anchoring sentences through its verbs.

Alternatively, phrase-structure representations, notably spearheaded by Noam Chomsky,[27] use constituency relations. In contrast with dependency trees, each 'constituent' (or, individual element) in a sentence is headed by its own phrasal node. Subsequently, purely binary branching can occur. The elegance of these representations is that they work generatively. That is, even a small selection of rules can produce a wide variety of structures found across natural language. Furthermore, constituency embraces analysis of underlying structure and transformations, accounting for numerous phenomena such as subject-verb inversion in interrogatives.[28] Phrase structure also permits a powerful structured analysis of syntactic relationships.[29]

[24] Oliver Wendell Holmes, *The Common Law* Lecture I: Early Forms of Liability (Project Gutenberg eBook, 2000), available at: https://www.gutenberg.org/files/2449/2449-h/2449-h.htm#link2H_4_0001.

[25] *Id.*

[26] *See* for example, Dan Simon, *A Third View of the Black Box: Cognitive Coherence in Legal Decision Making*, 71 U. CHI. L. REV. 511 (2004).

[27] Noam Chomsky, "Remarks on Nominalization," in R.A. Jacobs and P.S. Rosenbaum (eds.), *Readings in English Transformational Grammar* 184-221 (1970).

Semantic form traditionally involves the classical theory of concepts, otherwise known as definitionism or componential analysis. Here, semantic meaning is encapsulated as a combinatorial set of true/false statements, akin to a checklist of conditions. For example, *apple* might be composed of *+fruit, +green, +round.* Classical theory, therefore, considers the componential elements from which semantic meaning is formed, allowing for a systematic view on word-to-word relationships and validity.[30]

However, classical theory is often criticized for its failure to account for phenomena such as the subjectivity or typicality of definitions.[31] Ludwig Wittgenstein posited, through his analogy with 'family resemblance,' an underlying prototype theory of concepts; as opposed to a fixed set of composite definitions. The claim is that some concepts are regarded more 'typical' of a category than others. For example, a robin is a more prototypical bird than an emu or a penguin. Consequently, these observations must be factored into the linguistic system.[32]

What further complicates the matter is the incongruence between semantics and pragmatics: the former concerns language independent of real-world context, whereas the latter is hinged upon situational context. Essentially, pragmatics is the application of semantics within context.[33] Consider the phrase, "it's rather chilly in here." Semantically, the meaning of the phrase is perhaps that, according to the speaker, "there is a place X in which the temperature is lower than is comfortable." Given the knowledge that the phrase was taken from a dialogue between two individuals, the phrase pragmatically could mean "please close the window for me;"[34] the reason for the choice of phrasing is likely owed to courtesy. More importantly, this form of expression is indicative of the flexibility of language and its inseparability from context: context contributes to meaning.

While semantics concerns the inherent and invariant properties of words and their combinations, pragmatics progresses into the realm of context and implicatures. Consequently, pragmatics in the context of NLP is seen as problematic: expert systems do not have the ability to infer extended meaning from context. Interestingly, legal texts are often regarded as rather structural; perhaps even devoid of pragmatic content. Given the aforementioned premise, is legal language anchored exclusively in semantics? If so, how amenable is legal language to NLP analysis?

[28] Subject-verb inversion is the phenomenon whereby the verb is raised to a position in front of its subject, signalling an interrogative: "Have you seen my dog?". This raising is seen as a transformation.

[29] For example, the c-command relationship is easily identified, which is particularly useful when managing anaphora resolution through Government and Binding Theory (GBD). *See* Andrew Carnie, *Syntax: A Generative Introduction* (3rd ed. 2012).; and also Ray Jackendoff, *X Syntax: A Study of Phrase Structure* (1977).

[30] For further details: the classical theory of concepts presents a deconstructive view of meaning (semantics). By breaking words down into sets of necessary and sufficient conditions from a set of meta-concepts, we view their 'true' definition and form comparisons. For example, *bachelor* and *husband* suggest a commonality of *+male* but a distinction in the condition of *±married* (*-married* in the former and *+married* in the latter). *See* Eric Margolis and Stephen Laurence, *The Blackwell Guide to Philosophy of Mind* Concepts 190-213 (2003).

[31] *See* Ludwig Wittgenstein, *Philosophical Investigations* (2nd ed. 1953).

[32] Eleanor Rosch and Carolyn B. Mervis, *Family resemblances: Studies in the internal structure of categorie*s, 7 COGNITIVE PSYCHOLOGY 573 (1975).

[33] Keith Allan and Kasia M. Jaszczolt (eds.), *The Cambridge Handbook of Pragmatics* (2012). *See* also, Kasia Jaszczolt, *Semantics and pragmatics: Meaning in language and discourse* (2002).

[34] This is largely in line with discussions on conversational implicatures. *See* for example Henry E. Smith, *Modularity in Contracts: Boilerplate and Information Flow*, 104 MICH. L. REV. 1175, 1205 (2006).

### *c. Technological Staging: AI and Law*

Evidently, inspiration from linguistics is not novel as "law has language at its core."[35] Christopher Markou and Simon Deakin point to the breakthroughs in NLP that have contributed to the emergence of 'Legal Technology.' They identify the pressure points at which computability falls short; where the legal system is incompatible with computer science.

In fact, they cleverly evoke Chomskyan and rationalist approaches to designing "hard-coded rules for capturing human knowledge."[36] Chomsky's work stirred further developments in NLP, eventually powering advances in machine translation and speech recognition. These advances, undoubtedly, were enabled by Deep Learning[37] models that were able to abstract and build representations of human language. Albeit the significant leaps brought on by such technologies, the threat discussed by Markou and Deakin stems from an underlying anxiety against "the epistemic and practical viability of using AI and Big Data to replicate core aspects and processes of the legal system."[38]

Subsequently, their reimagining of a legal system – one predicated on a hyper-formalized method of reasoning[39] – warns of the conceivable incongruence with the current normative legal structure. Using employment status as a test case, their paper explores first similarities between legal processes and machine learning technology. They note two key parallels: (1) abstraction to conceptual categories; and (2) error correction and dynamic adjustment.

Nevertheless, their thesis, or claim of divergent paths, is the quality of reflexivity[40] in legal knowledge. That is, legal categories both shape and are shaped by the "social forms to which they relate."[41] In other words, the existence of such categories is dependent on the force of law;[42] that there is continual reference between the law and its socially complex environment. The law cannot be divorced from its societal embedding. As a result, the law could never be descriptive, but rather 'naturally' prescriptive. Markou and Deakin, therefore, identify a fundamental philosophical mismatch as opposed to a structural, process-oriented incongruity. Their conclusions underline legal reasoning as beyond the straightforward application of rules to facts. Adjudication is a means of "resolving political issues."[43] For Markou and Deakin, there is no exact science to judicial decisions "because of the unavoidable incompleteness of rules in the face of social complexity."[44] Judgments could only 'approximate' from historical precedent. Translation of legal categories into mathematical function is, thus, not possible since the flexibility and contestability of natural language cannot be completely captured by algorithm.[45]

[35] Christopher Markou and Simon Deakin, *Ex Machina Lex: The Limits of Legal Computability*, Working Paper (2019), available at SSRN: https://ssrn.com/abstract=3407856. *See* also Frank E. Cooper, *Effective Legal Writing* (1953) and his introduction with Law is Language; and "...the central place of language in law" described in Frank Pasquale, *The Substance of Poetic Procedure: Law & Humanity in the Work of Lawrence Joseph*, 32 LAW & LITERATURE 1, 31 (2020).

[36] *Id. See* also cited reference, E Brill and RJ Mooney, *Empirical Natural Language Processing*, 18 AI MAGAZINE 4 (1997).

[37] Deep Learning is a subset of machine learning that involves artificial neural networks and the assigning of numerical weights on input variables. *See* a further explanation in Markou and Deakin, *supra* 35 at 10-12.

[38] *Id.* at 16.

[39] *Id.*

[40] Markou and Deakin reference Geoffrey Samuel's discussion of the perception, construction and deconstruction of fact. See *id.* at 29. *See* also Geoffrey Samuel, *Epistemology and Method in Law* (2003).

[41] *Id.*

[42] The 'force of law' refers to HLA Hart's argument that the power of legal institutions and the laws created by such institutions exist in virtue of a rule of recognition implicit in the practice of judges. *See* Gerald J. Postema, *Implicit Law*, 13 LAW AND PHILOSOPHY 361 (1994).

Holmes's paradox resurfaces. Holmes notes, to "attempt to deduce the corpus from *a priori* postulates, or fall into the humbler error of supposing the science of the law reside[s] in the *elegantia juris*, or logical cohesion of part with part"[46] mistakenly interprets law as systemically formalistic. While the issues identified by Markou and Deakin are undeniably significant, their arguments rely on the premise of a systems replication. That is, they warn of the project to replace entirely juridical reasoning with machine learning. Accordingly, there are sweeping inferences on the incompatibility of AI and law, bringing to light only one side of Holmes's paradox: the law is syllogistic in form.

Yet, there may be merit to an analysis at a micro-level. Programming languages may be able to perform the demands called upon for the functioning of society. Acknowledging that language is both constitutive of law and capable of realizing foundational rule of law principles, we again reassess the translation of natural language to computer code. The law hinges on complicated social and political relationships;[47] and more importantly, metaphors that require latent understanding of temporal societal constructs.[48] This suggests there may be a space to regard AI as complementary,[49] rather than substitutive, of legal actors. The key is to employ the proper language game.[50]

For Lawrence Lessig, the conceptualization of code as law is not novel. Instead, he has drawn attention to code as a form of control in the 'cyberspace;' that "code writers are increasingly lawmakers."[51] The difficulty, of course, is defining the parameters of cyberspace. In this case, this may be the entirety of the legal sphere. To return to Markou and Deakin, their arguments repeatedly point to the model of 'legal singularity.'[52] 'Legal singularity' draws from an association of the law as precise, predictable, and certain in its function.[53] The complexity of developments in machine learning for law suggests that legal singularity could be achievable.

[43] Markou and Deakin, *supra* 35 at 30.

[44] *Id.* For further discussion on the 'incompleteness' and indeterminacy of the law, *see* Katherina Pistor and Chenggang Xu, *Incomplete* Law, 35 NYU J. INT'L L. & POL. 931 (2003).

[45] *Id.* at 33.

[46] Holmes, *supra* 24.

[47] Frank Pasquale, *A Rule of Persons, Not Machines: The Limits of Legal Automation*, 87 GEO. WASH. L. REV. 2, 6 (2019).

[48] Neil M. Richards and William D. Smart, "How should the law think about robots?" in Ryan Calo et al, eds, *Robot Law* 16-18 (2016).

[49] *See* for example Pasquale, *supra* 47.

[50] This is interpreted under the framework put forward by Wittgenstein. Wittgenstein regarded language as a form of life, and linguistic expression as constructive of its being. Conceivably, language could be no more than a list of orders and classifications. In abiding by the rules of association—or, to play the game—is to accept the inherent authority of its practice. *See* Wittgenstein, *supra* 31 at 11.

[51] Lawrence Lessig, *Code 2.0* 79 (2nd ed. 2006).

In a vibrant thought experiment, Casey and Niblett suggest that existing legal forms will become irrelevant as machines enable the development of a new type of law: the micro-directive. The micro-directive is conceptually a new linguistic form, offering "clear instruction to a citizen on how to comply with the law."[54] In this futuristic construct, lawmakers would only be required to set general policy objectives. Machines would bear the responsibility to examine its application in all possible contexts, creating a depository of legal rules that best achieve such an objective. The legal rules generated would then be converted into micro-directives that subsequently regulate how actors should comply with the law.

Imagining the legal order as a system of micro-directives, the law finds itself drawn to a linguistic structuralist framework; carrying forth the jurisprudential work of Kelsen and the "pure science of law."[55] Just as a norm expresses not what is, but what *ought* to be – given certain conditions – the micro-directive draws attention to the semiotics of legal argument. Like Kelsen's norms, the micro-directive rests on the principle of effectiveness. The legal order relies on the assumption of being efficacious, such that its citizens conduct themselves in pure conformity with it.[56] But, on what principle? The micro-directive rests on a 'law and economics'[57] framework of effectiveness. Seated within the technical authority of AI,[58] the micro-directive distorts the realities of legal reasoning by removing value judgments from the adjudication process. The presumption that machines are able to generate neutral sets of information, then translate such information into perfectly comprehensible instruction, is evidently misinformed. It stands on the premise that translation operates without interpretation. More importantly, it strategically excludes the actors involved in the translation; inadvertently, conferring the rule of law to code. The process of transforming a general standard to a micro-directive is, therefore, a process of subverting politics in its linguistic casing.

So, how then could code become the vehicle that shapes the law? In practice, the most obvious example is traffic laws and speed regulation. Traffic lights "communicate the content of a law to drivers at little cost and with great effect."[59] The traffic light is regarded as translating legal complexity to a simple command. Traffic lights are increasingly being equipped with algorithmic technology to reflect real-time traffic flow and, accordingly, adjust the timing of light changes.[60] Moreover, traffic lights may soon include sensors that could appropriately identify patterns of distress and types of vehicles to allow for expedited changes in the event of emergency.

[52] Benjamin Alarie, *The Path of the Law: Towards Legal Singularity*, 66 U. TORONTO L.J. 443, 445 (2016).

[53] *See* Freeman, *supra* 23.

[54] Anthony J. Casey & Anthony Niblett, *The Death of Rules and Standards*, Coase-Sandor Working Paper Series in Law and Economics No. 738 (2015).

[55] *See* for general theory on 'science' of law, Hans Kelsen, *Pure Theory of Law* (1967).

[56] Hans Kelsen, *What is Justice?* 268 (1957).

[57] Consider the argument for efficiency in common law rules (i.e. emergence of the economic loss 'rule') in Anthony Niblett et. al, *The Evolution of a Legal Rule*, 39 J. OF LEGAL STUDIES 325, 328-331 (2010).

[58] *See* discussion on algorithms as providing a convenient source of authority; trusting tasks to be controlled by technology and the delegation of responsibility. *See* Hannah Fry, *Hello World: Being Human in the Age of Algorithms* 16 (2018).

[59] Casey and Niblett, *supra* 54 at 18. *See* also, Sheila Jasanoff, *The Ethics of Invention: Technology and the Human Future* (2016).

For Casey and Niblett, predictive models provide the content of the law. Micro-directives would then communicate the legal treatment of the particular conundrum.[61] Legal actors would equally rely on such models to assess the acceptable plans of action for a particular diagnosis or factual circumstance. The micro-directive then reinvents the legal system, as legal language is eradicated and bears a different linguistic form.

Though at polar ends of the spectrum, both Markou and Deakin and Casey and Niblett depend on the same underlying assumption of a wholesale replacement of legal reasoning. This approach certainly raises significant metaphorical eyebrows on the broad impacts of AI in law. It, however, also avoids the nuances of the law that demand further analysis; in particular, the act of translation. Holmes described the "single germ multiplying and branching into products as different from each other as the flower from the root."[62] Thus, to make sense of the consequences of computational technology in law necessitates not an evaluation of the flower or the root, but the single germ.

Precision has often been argued as an essential component of legal language. Nonetheless, new factual circumstances create room for interpretation. How then could 'code-ification' occur to account for an ever-adaptive, and evolutionary, system? In the following section, the authors will outline the computational tools used in the translation process. More importantly, the authors will peel back the curtain behind translation; specifically, the decisions taken in the parsing of the legal judgments.

## II. METHODOLOGY

Prior literature on Deep Learning in legal text analytics traditionally discussed crafting knowledge bases to capture legal concepts and terminology.[63] Ilias Chalkidis and Dimitrios Kampas reflect on existing techniques, but push the envelope by building word embeddings[64] trained over a large body of legal documents; a corpora composed of legislation from the UK, EU, Canada, Australia, USA, and others.[65] Applying the Word2Vec model,[66] Chalkidis and Kampas's own model – aptly named

[60] *Id.* at 19.

[61] Casey and Niblett, *supra* 54 at 16.

[62] Holmes, *supra* 24.

[63] *See* for example Phong-Khac Do et al., *Legal Question Answering using Ranking SVM and Deep Convolutional Neural Network*, TENTH INTERNATIONAL WORKSHOP ON JURIS-INFORMATICS (2017), available at: https://arxiv.org/abs/1703.05320.

[64] Defined as a numeric representation of words whereby words with similar meanings would have similar representations. *See* Jason Brownlee, *What are Word Embeddings for Text?*, Deep Learning for Natural Language Processing (October 11, 2017), available at: https://machinelearningmastery.com/what-are-word-embeddings/.

[65] Ilias Chalkidis and Dimitrios Kampas, *Deep Learning in law: early adaptation and legal word embeddings trained on large corpora*, 27 ARTIFICIAL INTELLIGENCE AND LAW 171 (2018).

[66] A statistical model first introduced in 2013 that described two different algorithms used to process text into vectors. It is an example of the transformation of words to numeric form. *See* Mikolov et al., *Distributed representations of*

Law2Vec – offer a pre-trained set of legal word embeddings. Broadly, the process involves translating legal text to numeric form in order to calculate the relationships between legal terms. The calculation represents the probabilistic likelihood of one term appearing synonymous in the presence of the other. The main assumption is that "similar words tend to co-occur in similar contexts."[67]

Below is a table of a selected 20 words and their associated terms identified by the model:

| | |
|---|---|
| article | convention, section, articles, clause, provisions |
| act | statute, provision, mccarranferguson, irca, tvpa |
| action | suit, actions, lawsuit, claim, proceeding |
| crime | offense, murder, crimes, felony, violent |
| felony | offense, misdemeanor, felonies, offenses, convicted |
| punishment | penalty, punishments, sentencing, sentence, imprisonment |
| security | social, health, administration, retirement |
| fraud | fraudulent, theft, deceit, misrepresentation, bribery |
| privacy | confidentiality, communications, liberty, freedom, freedoms |
| intellectual | copyrights, patents, copyright, trademark, wipo |
| terrorism | terrorist, trafficking, counter-terrorism, violent, laundering |
| immigrant | immigrants, nonquota, alien, asylum, citizenship |
| illegal | unlawful, corrupt, improper, illicit, fraudulent |
| drugs | drug, narcotic, addicts, psychotropic, medicines |
| appeal | appeals, review, hearing, appellate, appealed |
| abuse | violence, sexual, self-destructive, assault, mistreatment |
| alcohol | liquor, spirits, intoxicating, beer, vinous |
| complaint | grievance, allegations, allegation, complaints, counterclaim |
| indictment | conviction, summary, imprisonment, indictable, triable |
| motion | motions, petition, dismiss, leave, cross-motion |

Table 1 Sample Legal Word Embeddings (Chalkidis and Kampas, *supra* 65 at 176)

This is undoubtedly remarkable. The associations made between the identified legal terms are indicative of the competence of machine learning algorithms for the analysis of complicated legal texts. Most fascinating perhaps are the terms associated with the word 'immigrant' found by the algorithm. Beyond locating synonyms, the terms deemed as similar reveal the latent politics of labelling that have classified immigrants as akin to aliens. Nevertheless, Chalkidis and Kampas offer only a limited perspective on legal concepts. The terms marked as 'legal' provide a scope of the law that does not consider the inherent interpretative exercise performed in adjudication. The act of legal reasoning is not represented. While Chalkidis and Kampas tease at the possibility of translation, the issue rather is arriving at the association. Chalkidis and Kampas could only bring to light the calculated similarities between legal terms; but they do not unpack *how* the similarity came about. In other words, the underlying process of deriving meaning is never exposed.

Moreover, the selection of terms deemed 'legal' are rather shallow. They are suggestive of a legal vocabulary, but do not probe at the function of these words. The authors, therefore, take inspiration from literature outside of the legal realm, focusing on the mechanics of linguistic reasoning and the adjudicative process.

*words and phrases and their compositionality*, PROCEEDINGS OF THE 26TH INTERNATIONAL CONFERENCE ON NEURAL INFORMATION PROCESSING SYSTEMS (2013).

[67] Chalkidis and Kampas, *supra* 65 at 173.

### a. *Technical Inspiration*

As an introductory note on method, Markou and Deakin have helpfully outlined NLP technologies that have set the sail on current applications of AI-based innovations.[68] NLP is a combined scientific and engineering exercise, applying "cognitive dimensions of…natural language" to "practical applications…[of] interactions between computer and human languages."[69] For the intentions of the paper, the focus will be on natural language in written form; otherwise, text. Not only is it the form in which law most typically resides, text is also the observable component of language that exists in symbolic form.[70] Interestingly, mathematics– or to recall, the mental alphabets of Leibniz and Boole – is described as *the* symbolic language.[71] It follows that translation is most feasibly comparable where both 'languages' are in a similar state.

In order for natural language text to be 'primed' for translation, the authors applied an approach first introduced in the sphere of bioinformatics. In 2006, Fundel et. al. developed RelEx, or the relation extraction of free text, to better understand the interactions between genes and proteins marked by existing biomedical publications. RelEx relies on natural language *preprocessing*, "producing dependency parse trees and applying a small number of simple rules to these trees."[72] RelEx extracts qualified relations from natural language text by first breaking down sentences into component words (tokens), then uses a parser[73] to create syntactic dependency trees. These dependency trees are then leveraged from group tokens into 'noun-phrase' chunks.[74] Qualified relations are observed based on rules applied to dependency trees and their original sentences; which are then subjected to 'filtering.'[75] These rules would draw paths that connect known proteins that interact with one another.

Analogously, the approach used in the RelEx paper will be applied to the current analysis of legal judgments. In addition to noun-phrases, sentences are deconstructed into the basic semantic building blocks of the English language;[76] otherwise, subject-verb-object (SVO) triplets. Sentences selected from each judgment are chosen based on their significance to the outcome of the judicial decision.

[68] Markou and Deakin, *supra* 35 at 11.

[69] *Id.*

[70] *Id.* at 12.

[71] *See* literature: Ladislav Rieger, *Algebraic Methods in Mathematical Logic* The Language of Mathematics and its Symbolization 25-37 (1967); Uttam Kharde, *The Symbolic Language of Mathematics*, 1 THE EXPLORER: A MULTIDISCIPLINARY JOURNAL OF RESEARCH 117-118 (2016); and Daniel Silver, *The New Language of Mathematics*, 105 AMERICAN SCIENTIST 364 (2017), available online: https://www.americanscientist.org/article/the-new-language-of-mathematics.

[72] Katrin Fundel, Robert Küffner, and Ralf Zimmer, *RelEx – Relation extraction using dependency parse trees*, 23 BIOINFORMATICS 365 (2006).

[73] Defined as a software that transforms data into structures.

[74] Defined as one or more nouns and their subordinate adjectives. *See* Fundel et. al, *supra* 72.

[75] *Id.* at 366.

[76] In other languages, a finite verb can occur without an overt pronominal subject. This is known as the null-subject, or pro-drop, parameter. The English language lends itself especially well to this approach due to the absence of this parameter. Furthermore, English generally does not allow zero copula forms (cf. Russian "я свободен" ("I [am] free")); this is also conducive to verb anchored SVO triplets in the dependency framework.

These sentences are subsequently scanned for the presence of SVO triplets. Markers are then assigned to each individual sentence based on equivalency, in order to then form connections between phrases.

Referring back to the aforementioned linguistic models, applying the RelEx method necessarily depends on a preference to dependency syntax and the classical theory of concepts (definitionism). Nevertheless, the authors argue that the mapping of each SVO component in reference to its neighboring components helps compensate the pitfalls involved with the multiplex nuances of word usage. By working with context, the analysis will extend beyond the realm of prototype theory,[77] which struggles to explain properties arising from context and pragmatic inference.[78] The graphing of the SVO triplets acknowledges context,[79] thereby becoming an integral part of the overall analysis. This method overlaps with ideas addressed in cognitive linguistics, such as the theory-theory of concepts, that heavily relies on role and context. Furthermore, employing sets of meta-concepts, along with graphed contextual relations, provides an analogy of traversing the semantic and pragmatic layers of language.

The project is, therefore, guided by three key tools: (1) Python; (2) spaCy; and (3) Neo4j. The first is the formal scripting language used to write the translation algorithm. Python was chosen for its known flexibility and general use.[80] Python also adapts in a number of design spaces, namely for tasks that are structural and reflective. spaCy is the chosen open source[81] library for NLP. spaCy is the primary software used to help parse sentences from legal texts to dependency trees; then to organize the components into categories.

The decision to use spaCy, as opposed to other NLP packages available in Python, is its ease of use, configurability, speed, and existing models pre-trained on a generalized data (e.g. articles, comments, blogs, etc.).[82] While intuitively NLP programs, such as LexNLP, were considered, the current test case poses a different challenge. LexNLP, for example, works with legal texts that are rather structured (i.e. contracts).[83] Therefore, LexNLP is trained at the document and clause level; thereby capable of extracting and classifying clauses as opposed to semantic content. The authors acknowledge that there are certainly merits to LexNLP. The greatest advantage being its models are pre-trained on U.S. legal texts. Nevertheless, spaCy offers much more functionality and flexibility given the breadth of subject matter found in the training data. By way of analogy, the choice may be akin to choosing between an oyster knife and a Swiss army knife when asked to descale a bass. The oyster knife is specialized but has its practical limits. In contrast, the Swiss army knife – emblematic of versatility – may offer more options and space for creativity when handling intricate tasks.

[77] Rosch and Mervis, *supra* 32.

[78] Jerry Fodor and Ernest Lepore, *The red herring and the pet fish: Why concepts still can't be prototypes*, 58 COGNITION 253 (1996).

[79] Defined here as other surrounding SVO elements.

[80] For further information on Python and developer knowledge, *see* Python, https://www.python.org/doc/.

[81] Open source is defined as software that is available for anyone to inspect, modify, and enhance. spaCy operates under a MIT license. This form of license is a permissive software license with the sole restriction that the original copyright and license notice be included in any future copies of the software. See *What is open source?*, opensource.com, https://opensource.com/resources/what-open-source. *See* also *The MIT License*, Open Source Initiative, https://opensource.org/licenses/MIT.

[82] For further details, see spaCy's technical documentation, available at: https://github.com/explosion/spacy-models/releases//tag/en_core_web_lg-2.2.5.

[83] *About LexNLP*, LexNLP, https://lexpredict-lexnlp.readthedocs.io/en/latest/about.html.

Finally, Neo4j is a graph database management system designed to store and process data in the form of nodes and relations.[84] The system helps classify the entities and the semantically relevant connections between such entities. Graph databases are commonly used for intermediate representation (IR). Known as the "stepping stone from what the programmer wrote to what the machine understands,"[85] IR is an object-oriented structure that, in its final form, stores all information required to execute a specified program.[86] IRs facilitate translations from natural language to machine code, bridging semantic gaps and behaving as the 'middleman' between syntactic forms. The graph database is also ideal for modelling dependency trees and object-oriented phenomena, such as inheritance. Put together, the authors attempt to advance the techniques inspired by RelEx for the translation of legal language to numeric form.

### b. *Risky Business: Case Selection*

The initial test cases selected for the POC are not arbitrary. The authors have strategically chosen cases that all follow a similar premise: what is the meaning of "use" applied to a firearm? Importantly, the cases belong to an alleged lineage; the application of precedent and consistency in legal adjudication.

In 1993, the Supreme Court of the United States (Court) was asked to rule on the definition of "use" in *Smith v. United States*. The petitioner, John Angus Smith, had offered to trade his gun in exchange for cocaine. He was subsequently charged with numerous firearm and drug trafficking offenses. This included using a firearm "during and in relation to" a drug trafficking crime, as stipulated under statute 18 U.S.C.§924(c)(1).[87] The Court held that the trading of a firearm constitutes "use" within the meaning of the statute. There are two remarkable notes to this case. First, the Court interprets the meaning of use rather broadly, particularly applying emphasis on the "everyday meaning and dictionary definitions" of use. Second, the interpretation is placed in the limited context of drug trafficking. The Court shifts away from a dictionary definition and, instead, emphasizes the furtherance of a crime as influential to the use.

In 1995, the Court was again asked to rule on the definition of use in *Bailey v. United States*. Similarly, the petitioners, Bailey and Robinson, were each convicted of drug offenses and of violating, none other than, 18 U.S.C.§924(c)(1).[88] The factual difference is the state of the firearm "during and in relation to" the drug-related offense. The Court was, therefore, asked to determine whether accessibility and proximity to the firearm was indicative of use. The Court held that the statute required "evidence sufficient to show an *active employment* of the firearm by the defendant, a use that makes the firearm an operative factor in relation to the predicate offense."[89] In *Bailey*, the Court then narrows the definition of use by including the element of "active employment." The Court provides a justification for its decision by referring to Smith and noting the ordinary definition of "use" in the active sense is "to avail oneself of."[90] Strikingly, the act of bartering falls within active employment, even though the gun was exchanged passively.

[84] For further information, *What is a Graph Database?,* Neo4j, https://neo4j.com/developer/graph-database/.

[85] Cliff Click and Michael Paleczny, *A Simple Graph-Based Intermediate Representation*, 1995 ACM SIGPLAN WORKSHOP ON INTERMEDIATE REPRESENTATIONS 35 (1995).

[86] *Id.*

[87] *Smith v. United States*, 508 U.S. 223 (1993).

[88] *Bailey v. United States*, 516 U.S. 137 (1995).

Coincidentally, a third case – three years later – had arisen, requesting the Court to rule on the definition of use under statute 18 U.S.C.§924(c)(1). However, *Muscarello v. United States* stretched beyond use and, instead, focused on "carries."[91] In *Muscarello*, enforcement officers had found guns in the petitioners' vehicles stored in a locked glove compartment and trunk respectively. The Court was, therefore, asked to determine whether that sufficiently fell within the definition of "carries." The Court ruled that carrying a firearm, in accordance with 18 U.S.C.§924(c)(1), "applies to a person who knowingly possesses and conveys firearms in a vehicle."[92] The Court again invokes "ordinary English," otherwise, basic meaning in dictionaries, to argue that carry is synonymous with conveys. Moreover, the Court again refers to *Smith*, but unlike *Bailey*, directs its reasoning to the *purpose* of the statute.[93] Notably, in all three cases, ordinary meaning was put forth as a dominant line of argumentation. Yet, the argument was always supplanted by intentions of Congress and the statute; that the purpose is to combat the "dangerous combination" of "drugs and guns."[94]

Funnily – perhaps to avoid a fourth case – Congress amended statute 18 U.S.C.§924(c)(1) to include "possess" in tandem with the phrase "in furtherance of any such crime;" thereby, accommodating the outcomes rendered in *Smith*, *Bailey*, and *Muscarello*. This then limited subsequent cases from arriving at the hands of the Court.[95] These cases were, therefore, carefully selected to illustrate that judicial decisions could bear the epistemic flavors of textualism with an underlying subtext of policy. Moreover, their similarity in factual circumstances allow for a stronger test of the underlying mechanisms of judicial reasoning and legal argumentation.

Again, the cases selected are not without limitations. In fact, they were cherry-picked to better demonstrate the subtleties of language and linguistics in law. Equally, the authors acknowledge that there are shortcomings to the project; namely, the importance of fact in law. Geoffrey Samuel states, "law arises out of fact."[96] That is, the legal effect of precedent extends so long as the material facts of the case are analogous. The project, however, does not account for the facts of the cases. Instead, they focus on the Court's specific arguments on the meaning of "use," accepting the facts as only peripheral to the exercise. The exclusion of facts may be problematic, given their significance to the nature of the common law system.[97] Still, the intentions of the paper are not to replicate judicial reasoning in common law. Fundamentally, the focus of the POC is translation, specifically an attempt to operationalize the migration of legal texts in natural language to algorithmic form.

[89] *Id.*

[90] *Id.*

[91] As a clarification, 18 U.S.C.§924(c)(1) involves both use and carries a firearm during and in relation to a drug trafficking crime.

[92] *Muscarello v. United States*, 524 U.S. 125 (1998).

[93] *Id.*

[94] *Smith*, *supra* 87 at 240. Also cited in *Muscarello*, *supra* 92.

[95] This is not to say no further cases were brought to courts involving the "use of a firearm" in a drug trafficking crime. This is only applicable to cases before the Supreme Court.

[96] Geoffrey Samuel, *A Short Introduction to the Common Law* 87 (2013).

## III. PRELIMINARY OBSERVATIONS

The inherent nature of interdisciplinary projects exposes the gaps between untraversed worlds. Between a data scientist, mathematician, linguist, and jurist, there are primarily two spheres of operation. One is derived from logic; and the other in humanities. Moreover, the disciplines speak different technical languages. Indubitably, there are clashes. Yet, the unifying mission to uncover 'meaning' has raised interesting perspectives on method and interpretation.

Consider the conversation between the linguist and computer scientist. The linguist struggles with a possible SVO markup for open clausal complements. The computer scientist suggests that it would fit 'cleanly' in the code if this were marked in the same manner as a clausal subject. The linguist is bewildered. In dependency linguistics, an open clausal complement is a clause without a subject. A clausal subject, on the other hand, is when a whole clause is itself a subject. What might be problematic with this type of equivalency?

This particular concern was contemplated within the framework of 'nested SVOs.' Complex sentences are composed of several clauses that carry condition and inherence. For example: adverbial phrases or subordinate clauses, that are themselves SVOs, act as modifiers to an overarching (superordinate) SVO. This became problematic when resolving the SVOs with one another; threatening a possible misalignment between their semantic and syntactic representation.

Another fascinating example came about when assessing the difference between the following two sentences:[98]

> "He shot the man with a gun."
>
> "He shot the man with a telescope."

For the human mind, the role of the object evidently differs between the sentences. In the former, the gun is indicative of the weapon used by the perpetrator. In the latter, the telescope is a qualifier of the victim, drawing a sharper image for the reader. This is owed to the cognitive association[99] between the object to the verb "shoot." But, what happens should the gun qualify "the man" in the first sentence? If so, not only does it change the meaning of the sentence, but, more importantly, it could affect the ultimate charge against the perpetrator. That is, the crime could be a difference between murder, manslaughter, or self-defense. The sentence alone cannot provide this depth of information required. Context and factual circumstances of the event are needed to determine how the sentence should be interpreted.

[97] Early origins of common law regarded it as a *customary* system of law, a body of practices observed by its players. *See* Vicki C. Jackson, *Constitutions as "Living Trees"? Comparative Constitutional Law and Interpretive Metaphors*, 75 FORDHAM L. REV. 921 (2006).

[98] It is important to note that the sentences are not taken from the judicial decisions but were conjectured in the process of completing the SVO markup.

[99] More specifically, the realm of psycholinguistics describes this association as top-down processing: the process through which knowledge and experience subconsciously influences interpretation of language. *See* Paul Warren, *Introducing Psycholinguistics* 137 (2013).

Interestingly, the data scientist and/or mathematician would approach the question by calculating the cosine similarity between the vector representations (word embeddings) of the verb and the object. Similar to the cognitive association performed in the human mind, the calculation determines the statistical probability[100] of the object appearing with the verb. The higher the frequency of both words co-occurring in the training corpus, the more likely the object is qualifying the verb. The cosine similarity can, therefore, be used as a numeric interpretation of how the object is employed given the verb in the sentence.

A third puzzle came in the form of homographs. Homographs, though identical orthographically, vary in meaning (though often distinguished in pronunciation). How then could a computer distinguish between record as a noun or record as a verb? The computer scientist notes that a distinction in the meta-concepts would resolve the problem. Meta-concepts, or metadata, are the elements outside of the SVO that describe the information being conveyed. This includes in what manner and how the sentence is being expressed. How important then is meta-data to the meaning of sentences?

This was again proposed as a possible resolution when encountering deictic expressions. Deictic words – such as 'this,' 'that,' 'here,' or 'there' – rely almost exclusively on context. Consider the sentence: "At issue *here* is not 'carries' at large, but 'carries a firearm' (emphasis added)."[101] What could 'here' mean? To the jurist, 'here' represented the material facts of the case, but to the linguist, it is a limited reference to the preceding sentences. To the mathematician or computer scientist, the word *here* represents a subjective concept for which a frame of reference and context serve to anchor it in reality.

These observations culminate to a greater question: what exactly constitutes as context? Meaning hinges on the knowledge of a "word by the company it keeps."[102] Should there be multiple interpretations of context, there are seemingly differing methods of arriving at 'meaning.'

At a glance, the SVO markups are products of conversations around these patterns of dependencies within sentences. Decisions were taken on how the sentences should be deconstructed to better articulate the interaction between subjects and objects with their verbs. Equally, an evaluation was made to separate meta-data from the basic SVO structures. Once the SVO markup was complete, it would form part of the training data for a decoder algorithm. The algorithm not only draws out the rules from the markup, but also other rules that the machine has gathered. This theoretically mirrors the concept of "reading between the lines." Finally, these rules are encoded for future documents in the graph created. The idea is that the markups identify only the more pertinent information in each sentence, while the algorithm detects any surrounding information.

[100] This is to reference the Word2Vec approach and transformer-based architectures that actively employ the surrounded words to mathematically derive context.

[101] *Muscarello* (Ginsburg, J., dissenting), *supra* 92 at 145.

[102] John Rupert Firth, *The Technique of Semantics*, 34 TRANS. PHILOS. SOC. 36 (1935).

The purpose is then to illustrate the connections and changes in the states of sentences found in the judicial decisions. In other words, it is the reconfiguration of sentences that are ostensibly void of structure, to their structurally dependent forms. In the following section, the authors articulate in detail the technical implementation of the project. They demonstrate that the translation of legal text to numeric form unravels the 'Black Box' of instinct[103] and disciplinary bounds. In the process of reducing sentences to SVO triplets, what is colloquially understood as intuition and knowledge-based expertise is revealed in a systematic form.

## IV. TECHNICAL IMPLEMENTATION

As discussed, there have been attempts at translating natural language to numeric form using various types of algorithms. To this day, success has primarily been achieved with the use of advanced statistical modelling techniques that depend on vast amounts of data. Leaning into these methods, the authors attempt to develop a new paradigm for natural language understanding; namely, one based on the core principles of Object-Oriented Design (OOD). The objective is to develop a preliminary model capable of ingesting a large amount of the data accurately, leaving the handling of outlier cases for a later stage of analysis.

The authors build on the ideas of Walter Daelemans and Koenraad De Smedt,[104] bridging concepts of OOD and linguistics. As the intention is not to be exhaustive, the table below broadly defines the analogies between OOD and linguistics that permit the translation of text into this form:

| **Object-Oriented Design** | **Concepts from Linguistics** |
|---|---|
| **Classes**<br>Blueprints (or prototypes) defining the characteristics and behaviors of Objects belonging to them | Hyponymy,[105] items contained in a set. Defining the prototype entities which allow objects to inherit any combination of single or multiple parent characteristics. |
| **Objects**<br>Singular manifestations of a Class | Noun-phrases and lexemes corresponding to singular entities and qualities (akin to individual definitions) represented by their lemma-form. |
| **Methods**<br>A defined interaction event between Abstractions in the program. Methods must be invoked in order for them to have a role. | Clauses (narrowed down to possible permutations of (S)V(O)) – interactions between semantic entities within the text.<br>The syntactic *subject* (semantic *agent*) is seen as the triggering *entity*, the (direct) *object* is the target of the interaction, and any additional *objects* behave as necessary inputs concerned in triggering the said interaction. The *verb* describes what happens during the interaction. |
| **Variables**<br>Placeholders for discrete information: values, Objects | Meronomy (declaring/assigning a placeholder for a part of the whole), meronymy (defining the content in the placeholder) – assigning parts of a whole. |

[103] Recall Simon, *supra* 26. *See* also R. George Wright, *The Role of Intuition in Judicial Decision making*, 42 HOUS. L. REV. 1381 (2005); and Chris Guthrie et. al, *Blinking on the Bench: How Judges Decide Cases*, Cornell Law Faculty Publications Paper 917 (2007), available at: https://scholarship.law.cornell.edu/cgi/viewcontent.cgi?article=1707&context=facpub.

[104] Walter Daelemans and Koenraad De Smelt, *Default Inheritance in an Object-Oriented Representation of Linguistic Categories*, 41 INT'L J. OF HUMAN COMPUTER STUDIES 149 (1994).

[105] To clarify, hyponymy describes the relationship of 'kind': if A is a type/kind of B, then A is a hyponym. In turn, meronymy is the relationship of 'parts' (also known as partonomy): if A is part of B, it is a meronym. For example, *table* and *chair* are hyponyms of *furniture,* whereas *wheels* and *doors* are meronyms of *car*. *See* Kate Kearns, *Semantics* (2000).

| | |
|---|---|
| **Abstraction**<br>The definition of Classes, Objects, Methods and Variables based on the task a program will solve. | Decoupling the signifier from the signified,[106] allowing for the open system nature of language and knowledge in general. |
| **Inheritance**<br>The passing on of characteristics and behaviors of a parent Abstraction onto its child | A multi-purpose mechanism allowing the modelling of linguistic phenomena, such as hyponymy, conducive to definitionism. |
| **Encapsulation**<br>The localization of characteristics and behaviors to a Class or Object | The phenomenon that allows for semantic parsing – localization of characteristics and behaviors to specific logical elements (entities) within a frame of reference. |
| **Polymorphism**<br>The ability to change any inherited characteristics and behaviors | Corresponding to the phenomena of polysemy and homographs, among others. Specifically, it allows any two entities within the same class to have different characteristics and behaviors represented by the same root word. |
| **Composition**<br>Arranging the interactions of Objects and Classes with one another; one of the aims of composition is to reduce code **redundancy** | Corresponding broadly to semantics – the arrangement, hierarchy and definition of communicative rules between the logical/semantic elements (perhaps equivalent to semes or sememes) in a text. This can allow for abstraction, improving **efficiency** in contextual assignment. |

As such, the grammatical structure of natural language is seminal to extracting its informational content. This would, in effect, permit a translation of 'meaning' to a form readily encodable in a programming language.

The complexity of legal concepts (i.e. the potential for multiplicity of meaning; or polysemy) called for technology that could cater to non-singularity. Consequently, the project attempts to strike a balance between definitionism and determinism by minimizing the pitfalls of both; the inefficiency and redundancy of definitionism against the brittleness of determinism. Ultimately, the goal is to secure efficient machine readability while upholding fundamental legal principles. The danger of leaning towards either the former or the latter is its adverse impact on the requirement for human intervention in the exercise of judicial reasoning. Should priority turn to definitionism, we risk creating a system that is far too complex and cumbersome to create any additional value for legal practitioners. Should priority turn to determinism, we risk creating a system that does not leave sufficient flexibility for ever-changing circumstances;[107] undermining existing legal structures.

Graph databases are amenable to generating highly interconnected webs of knowledge (knowledge-maps), optimizing analysis of relations between individual data points. Moreover, it accounts for issues of object composition, polymorphism, encapsulation, and inheritance; and enables the use of graph theory for creative analytical approaches on a larger scale. These ideas will return in the subsequent sections. Importantly, the graph works as the intermediary interface. It stores the input and analyzes the output of abstractions drawn from the developed algorithm.

[106] Referencing Ferdinand de Saussure and Jacques Derrida on semiotics. *See* Ferdinand de Saussure, *Course in General Linguistics* (Bloomsbury Revelations Ed. 2013); and Jacques Derrida, *Limited Inc.* (1988).

[107] Against the dismay of determinate expert systems, the authors are cognizant of judgments as temporally specific reflections of society; often, subject to influence by its sociopolitical environment. Notably, disruptions and shifts in society could (and often do) lead to reversal of judicial decisions. *See* for example the commentary by Kiel Brennan-Marquez and Stephen Henderson, *Artificial Intelligence and Role-Reversible Judgment*, 109 J. CRIM. L. & CRIMINOLOGY 137 (2019).

In the normal reading of texts, humans typically abstract in a sequential pattern; forming a 'world' within our own consciousness. Each subsequent phrase that speaks to the same topic enriches the details of this 'world,' reinforcing it with logical constraints and other abstractions.[108] This parallels a compiler reading a piece of high-level code, such as a Python script. The input works through layers of translation before arriving to a form comprehensible to the machine. Each stage serves to 'decompress' the knowledge built into the language by its designers. Eventually, the language is distilled down to its most granular level: a collection of binary code.[109] Phrases become a series of commands; either establishing a fact or describing an event or action.

The legal language is no different. It can be regarded as the sum effort of numerous iterations of layered abstractions rooted in social reality.[110] A legal document is the written manifestation of this process; conveying abstract legal concepts in a manner that is both syntactically sound and semantically meaningful in natural language.

One of the notable pitfalls of natural language is the underlying difference in contextual knowledge; whether it be prior experience or preconditioning. The existence of these differences manifests as "biases," which are then inherited in physical repositories, or artifacts.[111] Consequently, exposing context is often helpful in clarifying such 'repositories of legal knowledge.' For programmers, what is interpretable as context is the workings of reality outside the scope of a particular program. This could mean additional software may be used by developers when putting together a system (e.g. the importing of packages in Python). The addition of these packages extends the functionality of a program beyond its defined code. For the POC, the authors use a combination of pre-defined (i.e. spaCy's neural network models for recognizing dependencies and part-of-speech tags as well as Word2Vec converters) and newly trained estimators (i.e. detecting SVO triplets) to strengthen the model with metadata relevant to statements encountered in the dataset.

Below is a pictographic interpretation of the process:

[108] Warren, *supra* 99.

[109] To recall, this is an array of 0s or 1s to control transistors. It is the smallest unit of measure and often regarded in the logic form of an if-then statement.

[110] *See* for example Joseph Raz, *The Institutional Nature of Law*, 38 MODERN L. REV. 489 (1975); also, the difficulty of demarcating legal concepts in Joseph Raz, *Legal Principles and the Limits of Law*, 81 YALE L. J. 823 (1972).

[111] Langdon Winner, *Do Artifacts Have Politics?*, 109 DAEDALUS 121 (1980).

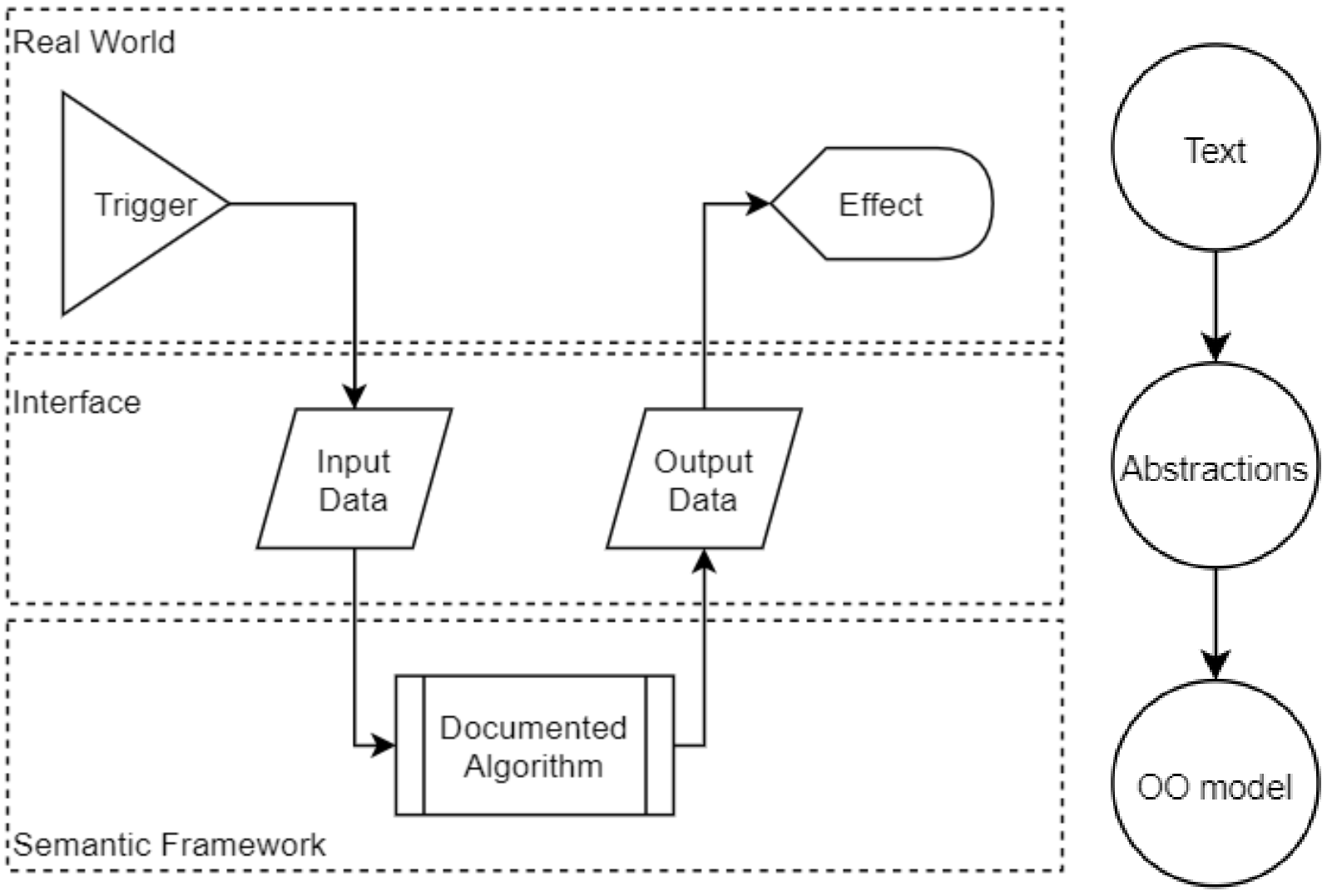

### *a. Defining Entities (Encapsulation)*

In building reference models of reality, entities are discrete units of existence. They act as mental placeholders to facilitate explanations of interactions within the model. Encapsulation is used to localize the characteristics and behavioral characteristic of each of these entities. The entities can be grouped into categories (classes), nested and (re-)arranged in an infinite number of ways. The importance is the architecture and its rules of performance; in other words, the process of defining entities of reference, their relations to one other, as well as their methods of interaction.

Consider the following sentence from *Bailey* as an informative example:

> "I *use* a gun to protect my house, but I've never had to *use* it."[112]

Disregarding first context, the sentence can be deconstructed into entities or methods. The entities, such as "I", "gun", "house," are encoded as nouns. The methods, such as "protect" and "use," are encoded as verbs.

Observably, the clause "I use a gun…" involves an actor ("I") that invokes an action ("use") on an object ("gun").

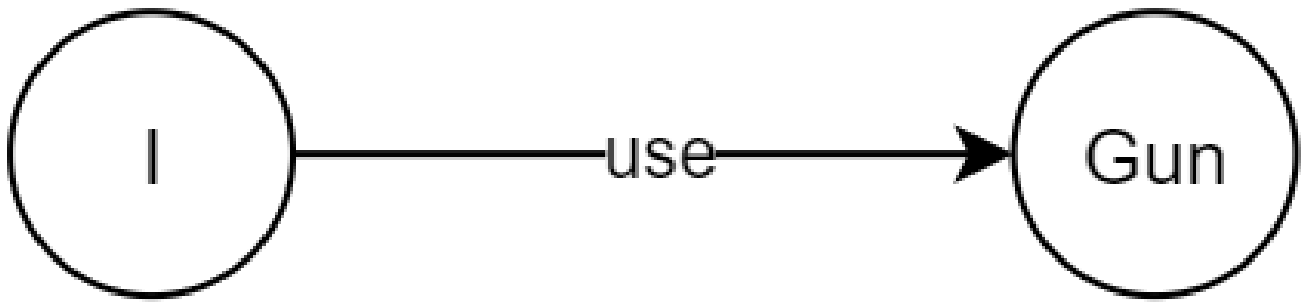

Applying the Object-Oriented approach of structuring code into classes and methods, the first phrase can be translated into the following schema:

[112] *Bailey*, *supra* 88 at 143.

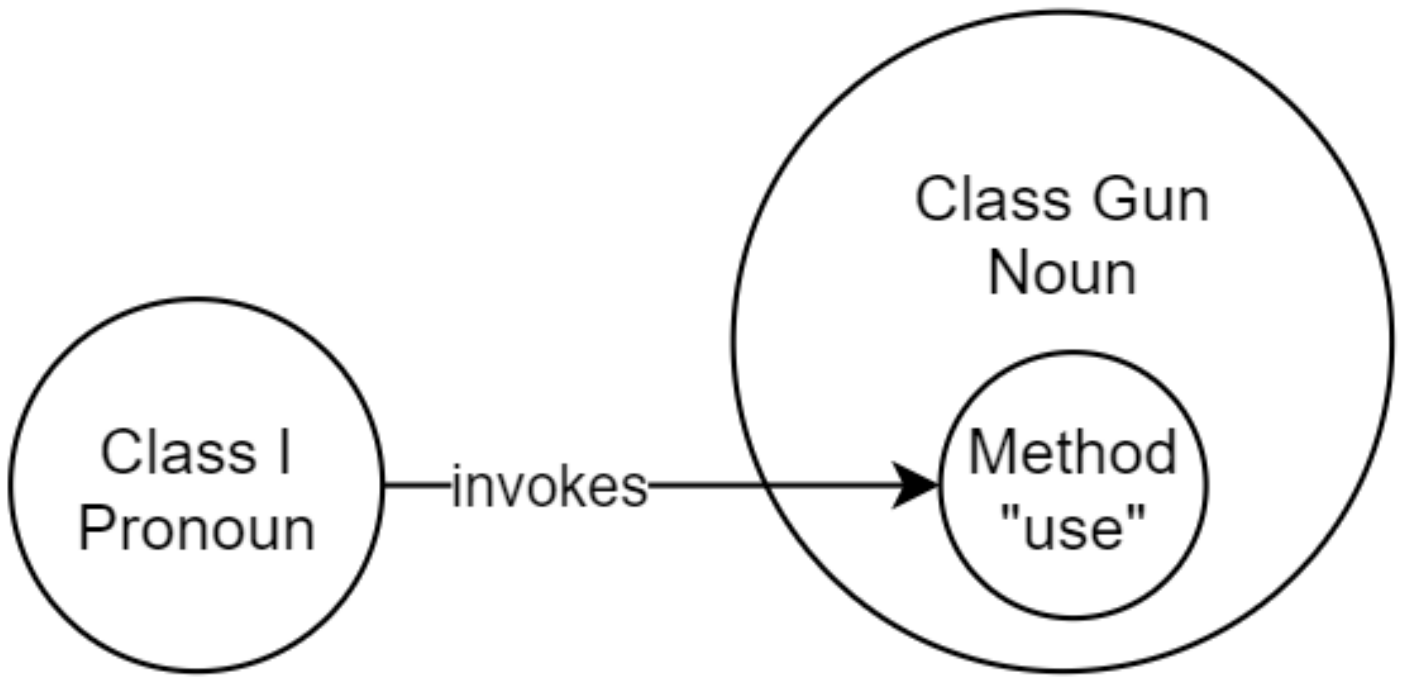

The components of the sentence become identifiable SVO triplets:

(1) the Subjects (invoking entity);
(2) the Objects (entity being acted on);
(3) the Verbs (method); and occasionally,
(4) the Prepositional Objects (additional entities describing the premise of the event/action).

The breakdown illustrates the framework on which the algorithm is built.

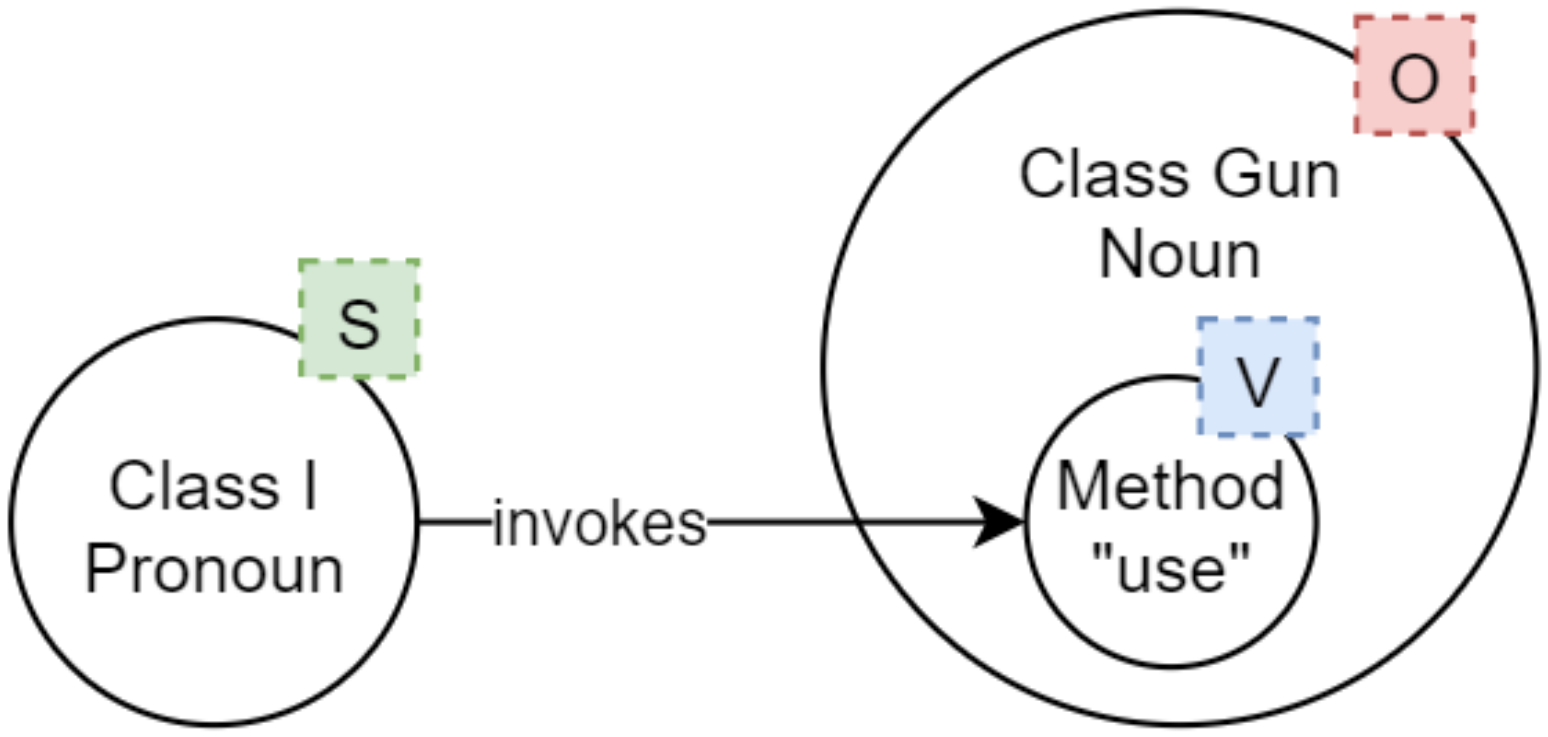

By extension of the example, subsequent phrases follow a similar breakdown, drawing connections between classes and their corresponding methods. This form of deconstruction also permits the nesting of concepts and additional logic tests along connections established.

### *b. Scaling Up (Composition)*

The process is akin to the first layer of translation; developing a pseudo-code script that represents a concept but expressible in a machine-readable language. The connections trace which class invoked the method "protect" on the class "house;" thereby deducing what "I" "use" to "protect" "house." As a result, such encoding does not require vast amounts of training data. Text is immediately translated to pseudo-code, without the need for external context.

The peripheral terms present in the sentence serve to indicate higher order concepts such as enumeration, negation, time, possession and pronoun assignment: "a", "never", "had to", "my", "it".

Their presence exists to modify the fundamental building blocks of the sentence - the nouns and verbs.

### *c. Creating the Knowledge Map (Natural Language Processing)*

Whereas the task of defining individual entities and methods is relatively straight-forward, creating a knowledge-map correspondent of the above schema requires the extraction of the semantic connections between them. By leveraging existing NLP tools,[113] such as spaCy, in conjunction with the authors' own SVO markups,[114] the authors were able to create a corpus to train a classifier capable of detecting SVO triplets and importing them to the graph.

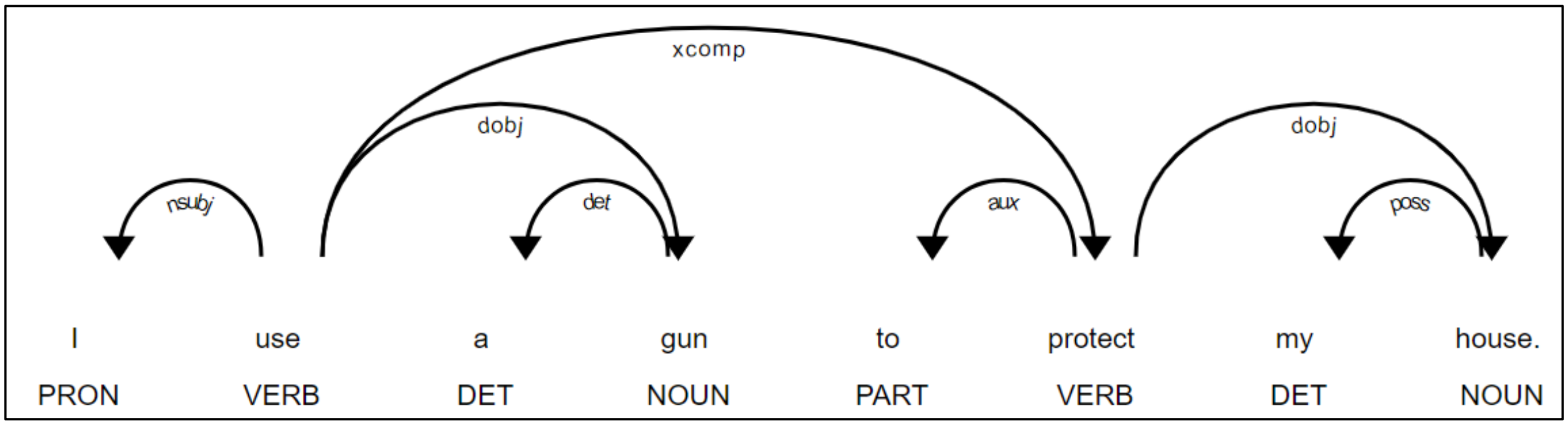

Figure A Sample Input to spaCy

The core strategy behind extracting SVO triplets lies in its linguistic deconstruction. The root of every sentence centers on the verb. Subjects ("nsubj") and objects (predicate, "dobj") are subordinate to verbs within the syntactic hierarchy. Therefore, in identifying the verbs of every sentence, the semantic connections are naturally found.

This method of text analysis has gained popularity with the advent of machine learning based models of NLP; trained on a sizable corpus of different expressions to perform the following tasks:

(a) Separating words from a string;
(b) Grouping the words into sentences;
(c) Assigning each word with a part-of-speech tag (Noun, Verb, Adverb, etc.); and
(d) Estimating each word's syntactic parent; thereby build a syntactic tree

This approach differs against other methods of semantic notation that rely solely on syntax; and less on the underlying pragmatics.[115]

[113] The authors do take note that even the most advanced language parsers are incapable of 100% accuracy. In analyzing the preliminary results, the authors have encountered a number of deficiencies owed to the dependency trees used. However, at this stage, the aim is again to capture a significant portion of the information within the text and leave outlier situations for the next stage of the project.

[114] Recall the RelEx method described in Section III. *See* Fundel et. al, *supra* 72.

[115] Recall subsection on Linguistic Influences and differences between dependency and constituency-based representations.

Between entity–method and SVO extraction, the data generated is sufficient to begin assembling together the knowledge-map. More importantly, the aforementioned process is derived entirely from the text itself. As a result, a defined cause-and-effect type algorithm is built, executable in full or in part, tested and queried. Additional metadata such as word embeddings, sentiment analysis and recognized named entities can provide supplementary information helpful for optimizing the knowledge-map and achieving a stronger understanding of the semantic content.

### *d. Building Character; Adding Context (Inheritance and Polymorphism)*

The project considers the transformation of legal texts to an Object-Oriented-like script; effectively using 'pseudo-code' to depict concepts embedded within the text. In natural language, multiplicity of meaning could occur when a single concept applies to several circumstances. Different conclusions can be drawn depending on the characteristics inheritable from a parent class. To clarify, this would include determining whether a "firearm" is within the same class as "gun." Similarly, other characteristics may include the methods or actions (verbs) invoked by a particular class. In object-oriented design, this phenomenon is known as polymorphism.

A core aspect of the translation to object-oriented form, as described in Daelemans and De Smedt's paper, is the assumption that subclasses 'inherit' the characteristics of the parent class by default; unless they are hard-coded otherwise.[116] In this case, characteristics and their behaviors are explicitly stated in the legal text. Consequently, if necessary and provided sufficient examples in the source text, as well as a threshold occurrence ratio, it will be possible to migrate certain characteristics up the inheritance hierarchy. Any such event can be signaled with a flag that the presence of this characteristic is an assumption with X percentage occurrence rate among child objects.

[116] Daelemans and De Smedt, *supra* 104.

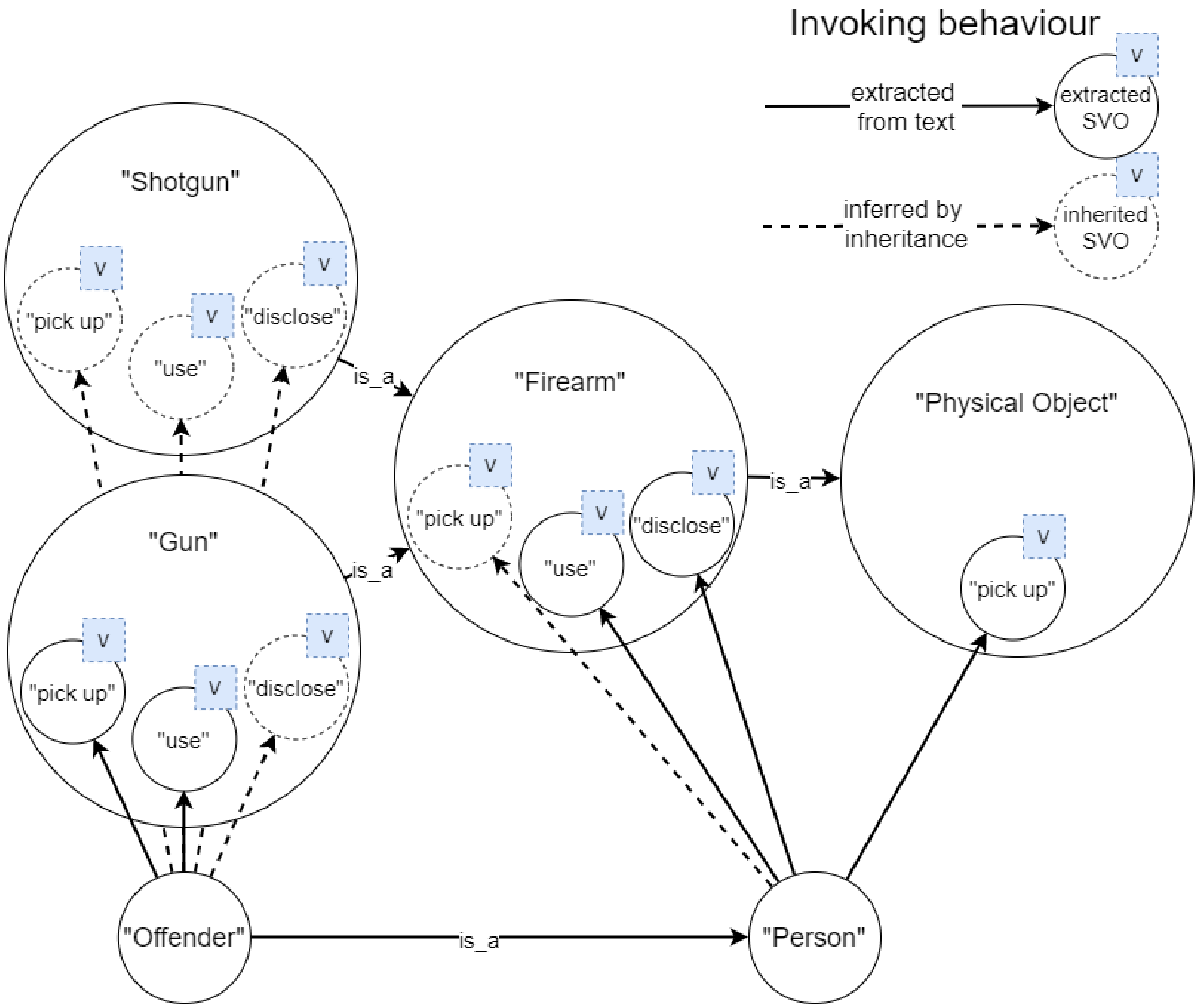

Figure B Illustrating Parent and Child Classes

When SVOs have an explicit subject and object, they can be loosely chained. However, the presence of subordinate clauses in the text necessitate nesting SVOs within one another. This exists in the pseudo-code as implicit causality. To then define the chain of causality, yet maintain the independence of each SVO, the root of a sentence must be identified. Drawing from the example, "I" must first "use a gun" in order to then "protect" "house". This suggests that "use" is the primary connection between the SVOs as one cannot exist without the other.

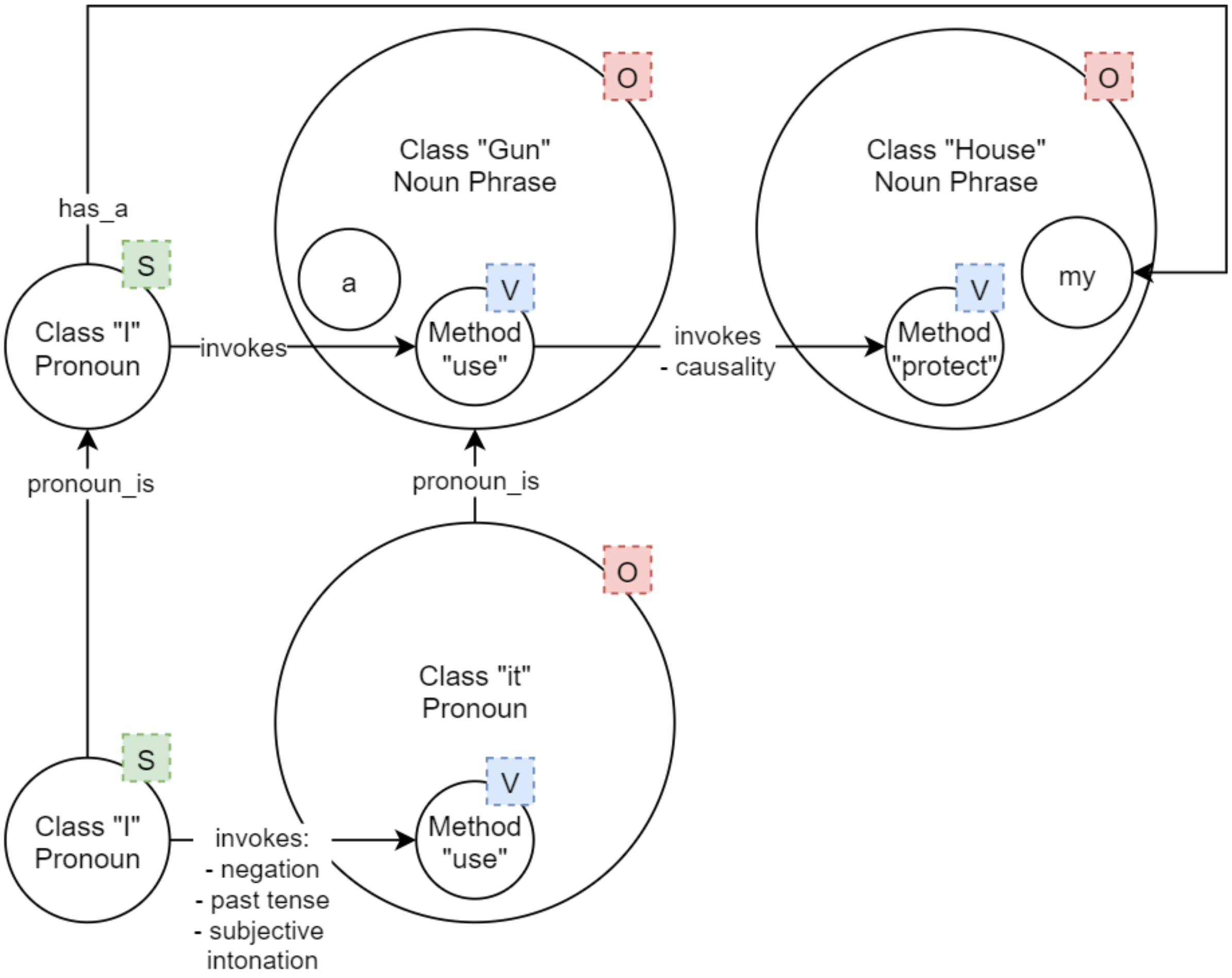

Figure C Illustrating causality between SVOs

Further classifications and qualifying characteristics may be important in a legal analysis. This information parallels the referencing of statutes and case law for prior interpretations of meaning. Various sources of law often create an environment for conflicting readings of a particular text. To tackle this problem, it is possible to assign an authority metric to each source; thereby establishing hierarchical structuring of the corpus. The structure behaves as a type of input when conducting an analysis, mirroring the hierarchy of legal sources.

## V. EARLY ACHIEVEMENTS AND FURTHER CONSIDERATIONS

### *a. By the word of the law*

Once the data was loaded into the graph, so began the stage of analysis. The primary way of interacting with the knowledge graph is the query function. Each query attempts to build one or more paths between two entities, with specific constraints along its path. This is the programming equivalent to writing tests for a piece of code. The knowledge graph is asked a question and returns a response that follows the reasoning of human observers. Once the knowledge graph has acquired sufficient data, the intention is to develop a user interface able to answer ‘legal’ questions posed by its users.

An invaluable tool used in this task is the Cypher query language. This language permits the formulation of queries based on the paths present within the data. The choice of constraints for each query will initially be hard-coded. Nevertheless, it is possible to then transfer the process to machine learning should sufficient data be gathered.

The idea behind this approach is to shift out of the standard statistically driven paradigm and allow the inference of logical conclusions from the text.

Consider a user query: "Describe the interactions involving a firearm."

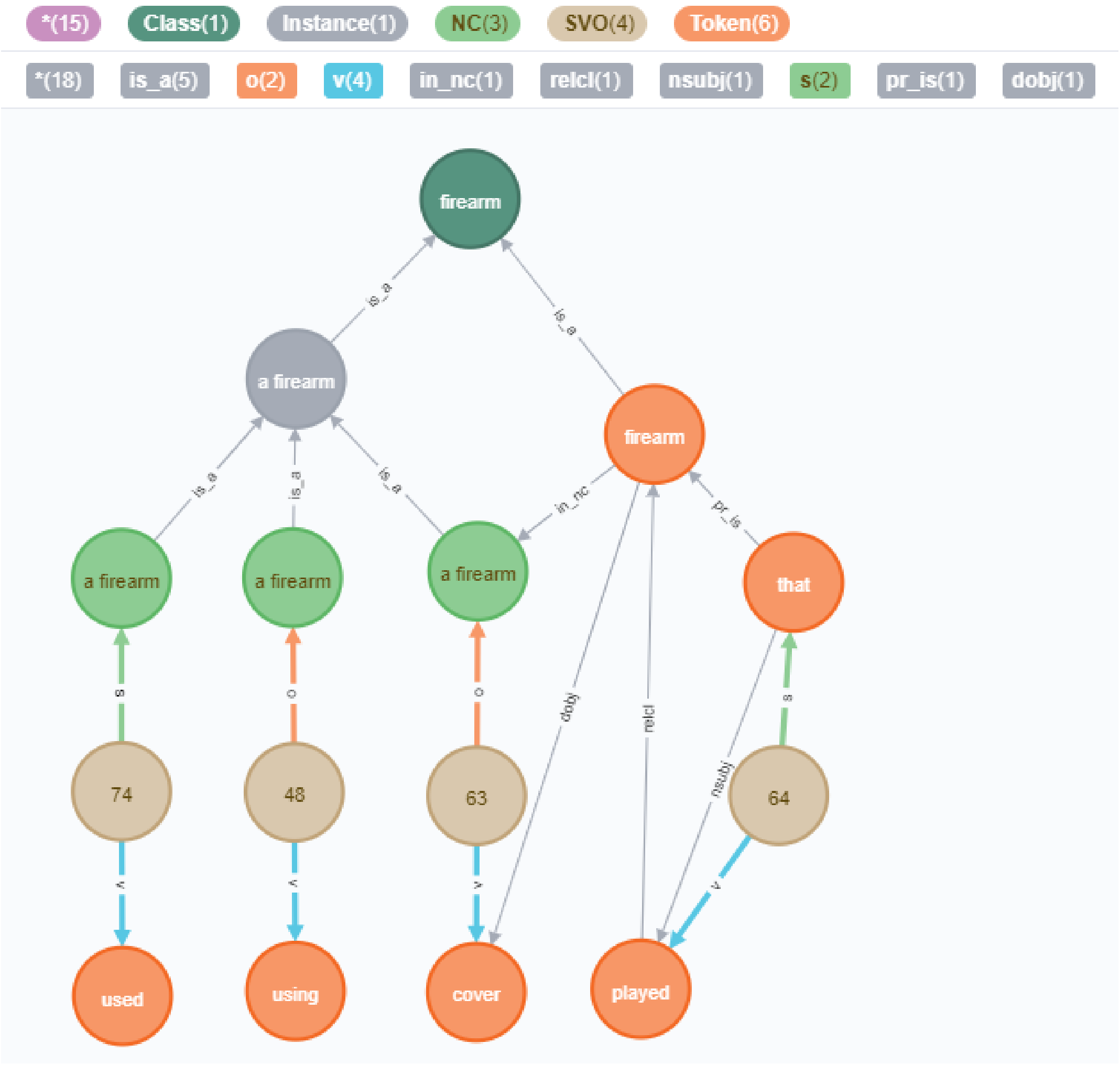

Figure D Sample Output from Neo4j Graph

With a user interface, the authors envisage that any question will be deconstructed in the same way as the training dataset. In this case, the algorithm should return the associations of entities and methods affiliated with "use" and "firearm." The interface will attempt to: (1) link the entities in the question, using the data in the graph; (2) gather any conditions and constraints along the way; and (3) return the relevant information as a series of possible paths taken within the graph, resulting in a list of phrases sorted by relevance (e.g. "use is active employment").[117] In effect, legal judgments are reconfigured into machine readable form to identify the meaning of the text. The graph acts to signpost legal actors towards definitions found in judicial decisions; thereby augmenting legal reasoning by leveraging the efficiency and power of computational analysis.

### *b. By the sixth sense*

On the other hand, there has been a latent understanding that intuition plays a role in the rendering of judicial decisions.[118] The techniques used in the authors' approach, in fact, account for instinct. The parsing of legal texts requires two types of algorithmic methods: (1) analytical; (2) and numerical.

The former serves to build a rigid structure from text and establish a hierarchy of semantic content on the basis of clearly defined criteria. This was demonstrable in the use of the graph database. The latter leverages the statistical modelling principles of neural networks. Similar to impulses attributable to intuition, the weight of each neuron in a neural network can be viewed as an abstract meta-concept; too complex to express tangibly.  A parallel can be drawn between the phenomenon of a "gut feeling" to the internals of a neural network, as trends embedded within a dataset are sorted into an array of codependent activation values. This means that any data present on the graph can be fed to customized machine learning algorithms to approximate human 'intuition.' Together, the authors could factor several forms of legal reasoning that often underlie judicial decisions.

### *c. Between implementation and effect*

To come full circle, the impact of translation has inadvertently exposed the logic of legal reasoning. Whether it is judicial intuition or syllogistic application, Holmes's paradox remains relevant. Words of legal text do, in fact, intrinsically embody meaning. The sphere of legal knowledge exists well within the sentences of judicial decisions. This is owed to the interpretation and conceptualization of precedent. The POC has observed that the use of precedent is not a procedural legal tool but a substantive one. Its application is to uphold the appearance of methodological consistency within the body of law. Yet, fundamentally, its use is to substantiate the authority of legal texts.

More importantly, precedent recognizably functions in an asymmetrical, as opposed to syllogistic, manner.[119] To recall, *Bailey* does not apply the plain meaning of 'active employment,' but constructs instead an alternative legal meaning to equate 'active' as "operative factor."[120] In other words, in accordance with *Smith* and *Bailey*, the use of a firearm includes bartering; and as such, the trading of a firearm is an 'operative' component to a drug-trafficking crime. These definitions are not logically deduced. Instead, they seek to reinforce a specific legal framing. Arguably, then, the use of precedent is not to follow past decisions, but to determine how to align with them. This was integral to incorporate in the graph, as the semantic content drew from legal taxonomy.

[117] *Bailey*, *supra* 88 at 137.

[118] Recall discussion on intuition in judicial decision making; *see* Wright and Guthrie, *supra* 103.

[119] Countering Holmes's description of the law as following syllogistically from existing precedents. *See* Holmes, *supra* 24.

[120] *Bailey*, *supra* 88 at 143.

The result of translating legal text in the manner described in Section 4 corroborates that legal language is self-referential and consistent. The law pushes outward by looking inward. In deconstructing legal judgments to its constituent components, the process of applying precedent evidently evolves: from syllogistic application to a framework of extraction.

### *d. On new terrain*

One of the aims of the research is to test the feasibility of expanding this approach; to translate natural language into code over a bigger corpus. When navigating the knowledge graph, the authors intend to apply elements of deontic logic[121] as an initial approach to the causal resolution of sequential events. Each SVO will be attributed variables corresponding to whether or not it is possible and/or necessary. This determines whether it can or should be invoked when resolving a query. Furthermore, the data can be used as input for graph-based algorithms when carrying out large-scale data analysis.

The next phase considers how far this model can extract the semantic content of judicial decisions. The application of the knowledge-map will be useful in this regard. Having laid the groundwork, the project will progress towards detecting logically conflicting statements, adding dimensions such as time and conditionality, as well as testing more nuanced relationships found in the text.

First, the technical implementation of the POC has alluded to the possibility of conflicting interpretations of legal text fostered by the array of legal sources. Alternatively, the presence of dissenting opinions could pose a similar issue (i.e. opposing views on a subject matter). This may appear on the graph as identical SVOs; one in the affirmative and the other negated. As all SVOs will be assigned an authority metric based on the source of law, it may be possible to include an additional marker indicating: (1) the opinion of the court; (2) dissent or concurring opinions. This would enable navigating any query along the SVO by the highest level of authority; while equally maintaining any 'lower level' paths that would offer informative insight.

Next, some SVOs denote conditions that invoke a parent SVO (e.g. phrases beginning with "if," "when," "whether," etc.). Such phrases are equivalent to tests that must be satisfied by a different set of nodes within the graph, providing answers to specific queries. By labelling these SVOs as conditional in the graph, it is possible to connect them as preconditions for the invocation of another SVO. For example, "if one 'carries a firearm,' then he or she knowingly possesses it."[122] The "if" statement denotes a logical test to determine whether the firearm is "knowingly possessed."

Owing to their subjective and relative nature, it remains undecided whether and how to group seemingly similar modifiers. Modifiers (i.e. adjectives and adverbs) used to qualify the root of each component in an SVO are currently treated as separate objects for the purpose of comparison. For the intentions of the project, modifiers can enable a connection between semantics and pragmatics embedded within the legal text. As the project uses legal text to generate an abstract frame of reference, the decision to treat modifiers as purely subjective results in a subjectively impartial point of view.

[121] A branch of symbolic logic influenced by modal notions. See *Deontic Logic*, Stanford Encyclopedia of Philosophy (April 21, 2020), https://plato.stanford.edu/entries/logic-deontic/.

[122] To paraphrase *Muscarello*, *supra* 92.

Consider the phrase "a fast dog might run alongside a slow car." The modifiers 'fast' and 'slow' can apply to the same quantity of an entity (e.g. degree of speed). It is only then sensible to compare/connect isolated relative terms directly within the bounds of their entity's relative frame of reference with regards to the same modifier. Having narrowed the scope, it becomes reasonable to connect purely subjective qualities (e.g. dog: "fast"→ is_not/faster_than → "slow").[123] As well, it is necessary to distinguish the subjective and objective qualities of any entity, with the subjective dependent on the objective. The establishment of this connection can serve to anchor subjective qualities from multiple sources within a common empirical frame of reference; rendering them comparable. Should the legal text provide sufficient information, the authors intend to create a mechanism to affix any subjective qualities to a quantifiable framework (e.g time,[124] position, etc.).

Last, it is possible to build predictive modules using graph theory that will infer correlations, trends, and common behaviors between the nodes and edges within the graph, as well as integrate external factors (e.g. socio-economic factors, business metrics, etc.). For instance, there may be interest in exploring the importance of certain SVOs in the graph relevant to their structural surroundings. The goal could be to analyze the patterns in the paths connecting SVOs. This is one area where a structural analysis, using graph theory, is helpful. Random walks[125], popular in machine learning, could analyze the latent representation of nodes within a graph. In this case, randomized paths initiated at a subset of nodes would be taken to collect sets of metadata used to infer trends within the structure of the graph. This has the potential to describe how the relationships between related SVOs evolve in legal discussion and could also illuminate underlying real-world effects.

Applications of such an approach could include:

- Evaluating the key logical paths characteristic of a precedent
- Highlighting areas of increased ambiguity within legal reasoning
- Estimating and counter-acting implicit bias
- Identifying phenomena such as feedback-loops and contradictions

## CONCLUDING REMARKS AND NEXT STEPS

The fundamental question asked by the project is whether meaning draws association from the language in which it is seated; that in changing the language, meaning will naturally be reconceptualized. The test to translate natural language to numeric form is not novel. In fact, it follows an ancestry of applying mathematical precision to legal expression. This paper has sought to experiment with the conversion of legal texts into algorithmic form. More importantly, the authors attempted to capture legal concepts and processes involved in legal reasoning. The deconstruction of natural language phrases to SVOs atomized sentences to their bare structures; forcibly exposing connections integral to the formation of concepts. As the authors aimed to reconcile syntax with semantics, structure became indistinguishable from content.

[123] Since the concept of speed can be applied to a number of contexts, it is then sensible to encapsulate it in its own class connected to the modifiers.

[124] For example, the time at which an event occurs plays a critical role in setting up the semantic network of facts within a knowledge graph. Each SVO will be marked with a temporal factor, both relative and absolute, which can later be compared for inconsistencies.

[125] A random walk is a set of randomized movements on our graph that yield a path through a set of nodes. For a detailed use case, *see* Bryan Perozzi et. al., *Deep Walk: Online Learning of Social Representations*, Proceedings of the 20th ACM SGKDD International Conference on Knowledge Discovery and Data (2014), available at: https://dl.acm.org/doi/10.1145/2623330.2623732

Inadvertently, the POC has demonstrated that, though form is seminal to the adjudicative exercise, the logic embedded within legal texts does not necessarily behave syllogistically. Instead, legal concepts appear to evolve sporadically. This sporadicity, however, is not synonymous to randomness. Rather, the development of the law draws from introspection and uses precedent to substantiate its authority. Teasing at Holmes's paradox, the law approaches consistency not in form, but in substance. As opposed to syllogistic application, meaning is found through a process of extraction.

The next phase of the project will bring forth a deeper breakdown of legal texts, focusing on higher levels of abstraction (i.e. trends latent in meta-concepts) and more complex grammatical resolutions found in natural language. From a broader perspective, the project has inspired the authors to advance towards a 'White Box' solution. The aim is to strengthen the understanding of legal texts, providing better roadmaps and signposting users towards more consistent interpretations of judicial decisions. It is an evolution of legal reasoning that heightens transparency and auditability by unpacking juridical truths and structuring intangible legal narratives. The result? Improving the quality of legal analysis and elevating accessibility to society.

As opposed to "grafting new technology onto old working practices,"[126] it is a new embodiment of precedent. It is a harnessing of the future through a preservation of the past. The integration of computational technology in law disrupts conventional legal mechanics, while maintaining the function of law. The authors anticipate a Bilbao effect, that the thoughtful marriage of old and new architecture sparks transformation.

[126] Referencing the distinction Susskind makes between automation and transformation. *See* Richard Susskind, *Online Courts and the Future of Justice* 34 (2019).